\documentclass[10pt]{article}
\usepackage{amssymb,amsmath}
\usepackage{amsfonts}
\usepackage{eufrak}
\usepackage{graphicx}

\begin{document}

\begin{center}
{\large {\bf Transverse Wave Propagation in Relativistic\\
Two-fluid Plasmas in de Sitter Space}}\\
\vspace{2cm} M. Atiqur Rahman\footnote{\it E-mail:
atirubd@yahoo.com} and M. Hossain Ali \footnote {\it The Abdus Salam
International Centre for Theoretical Physics, Strada Costiera 11,
34014 Trieste,
Italy. E-mail: mali@ictp.it, $m_-hossain_-ali_-bd@yahoo.com$ (Corresponding author).}\\
{\it Department of Applied Mathematics, University of Rajshahi ,
Rajshahi-6205, Bangladesh.}
\end{center}
\vspace{3cm} \centerline{\bf Abstract}
\baselineskip=18pt
\bigskip

We investigate transverse electromagnetic waves propagating in a
plasma in the de Sitter space. Using the $3+1$ formalism we derive
the relativistic two-fluid equations to take account of the effects
due to the horizon and describe the set of simultaneous linear
equations for the perturbations. We use a local approximation to
investigate the one-dimensional radial propagation of Alfv\'en and
high frequency electromagnetic waves and solve the dispersion
relation for these waves numerically.
\vspace{0.5cm}

\vfill

\newpage

\section{Introduction}\label{Intro}
In recent years there have been renewed interests in investigating
plasmas in curved spacetimes of general relativity; because, a
successful study of the waves and emissions from plasmas falling
into a compact body (e.g. black hole) will be of great value in
aiding the observational identification of black hole candidates.
The de Sitter (dS) space with positive cosmological constant has
properties similar to a black hole. We study two-fluid plasmas near
the horizon of the pure dS space.

Over the last few decades, physicists have a growing interest in dS
space. In the 1970s, the attention was due to the large symmetry
group of dS space, which made the field theory in dS space less
ambiguous than, for example, in the Schwarzschild spacetime.
Researches in the 1980s focused the role it played during
inflation--accelerated expansion in the very early universe. The
universe is currently asymptotic dS and approach a pure dS space.
Recent cosmological observations \cite{one,two,three,four,five,six}
suggest the possibility of existing a positive cosmological constant
($\Lambda>0$) in our universe and this possibility gives the
picture, among many others, of some features closely related to
black holes: the existence of cosmological event horizons. These
causal horizons exist even in the absence of matter, namely in empty
dS space, and hide all the events which are not accessible for
geodesic observers. In addition, the success of the ADS/CFT
correspondence \cite{seven,eight,nine,ten} has led to the intense
study of dS space in the context of the quantum gravity
\cite{eleven}. The attention has been to obtain an analogue of the
ADS/CFT correspondence in dS space, i.e. dS/CFT correspondence
\cite{twelve,thirteen,fourteen,fifteen,sixteen,seventeen,eighteen,nineteen,twenty,twenty
one,twenty two,twenty three,twenty four,twenty five, twenty six} in
the light of which there has been an extensive study of the
semiclassical aspects of dS and asymptotic dS spacetimes
\cite{twenty seven,twenty eight,twenty nine,thirty,thirty one,thirty
two,thirty three}. In view of these reasons, it may be of special
interest to investigate electromagnetic waves in a plasma in the dS
space.

Recently, Buzzi, et.al. \cite{thirty three,thirty four}, using the
3+1 formulation \cite{thirty five,thirty six,thirty seven,thirty
eight}, described a general relativistic version of two-fluid
formulation of plasma physics and investigated the nature of plasma
waves (transverse waves in \cite{thirty three}, and longitudinal
waves in \cite{thirty four}) near the horizon of the Schwarzschild
black hole. In this paper we apply the formalism of Buzzi et.al.
\cite{thirty three} to investigate the transverse electromagnetic
waves propagating in a plasma close to the (cosmological) event
horizon of the pure dS space.

This paper is organized as follows. In section \ref{sec2} we
summarize the 3+1 formulation of general relativity. In section
\ref{sec3} we describe the horizon plasma governing equations. In
section \ref{sec4} we consider one-dimensional wave propagation in
the radial $z$ (Rindler coordinate system) direction. We linearize
the equations in section \ref{sec5} by considering a small
perturbation to fields and fluid parameters. In section \ref{sec6}
we discuss the local or mean-field approximation for the lapse
function $\alpha$ and obtain a dispersion relation for the
transverse wave. In section \ref{sec7} we describe a procedure for
solving the dispersion relation numerically. In sections \ref{sec8}
we present our results. Finally, in section \ref{sec9} we give our
remarks. We use units $G=c=k_B=1$.

\section{3+1 Spacetime Formalism}\label{sec2}
In de sitter space, the simplest solution for the Einstein field
equations with $T_{\mu\nu}=0$ is written as
\begin{eqnarray}
ds^2&=&g_{\mu \nu }dx^\mu dx^\nu\nonumber\\
 &=&-\left(1-\frac{r^2}{\ell^2}\right)dt^2
 +\left(1-\frac{r^2}{\ell^2}\right)^{-1}dr^2+r^2d\Omega_2^2.\label{eq1}
\end{eqnarray}
Here, $\ell$ is the curvature radius of the dS space [i.e.,
$\Lambda=\frac{3}{\ell^2}$ is the positive cosmological constant],
$d\Omega_2^2$ represents a unit 2-sphere, and the nonangular
coordinates range according to $0\leq r\leq\ell$ and $-\infty\leq
t\leq\infty$. The boundary at $r=\ell$ describes a cosmological
horizon for an observer located at $r=0$.

An absolute three-dimensional space defined by the hypersurfaces of
constant universal time $t$ is described by the metric
\begin{equation}
ds^2=g_{ij}dx^idx^j
=\left(1-\frac{r^2}{\ell^2}\right)^{-1}dr^2+r^2d\Omega_2^2.\label{eq2}
\end{equation}
The indices $i$, $j$ range over 1, 2, 3 and refer to coordinates in
absolute space. The Fiducial Observers (FIDOs), the observers
remaining at rest with respect to this absolute space, measure their
proper time $\tau $ using clocks that they carry with them and make
local measurements of physical quantities. Then all their measured
quantities are defined as FIDO locally measured quantities and all
rates measured by them are measured using FIDO proper time. The
FIDOs use a local Cartesian coordinate system with unit basis
vectors tangent to the coordinate line
\begin{equation}
{\bf e}_{\hat
r}=\left(1-\frac{r^2}{\ell^2}\right)^{1/2}\frac{\partial }{\partial
r},\hspace{1cm}{\bf e}_{\hat \theta }=\frac{1}{r}\frac{\partial
}{\partial \theta },\hspace{1cm}{\bf e}_{\hat \varphi
}=\frac{1}{r\,{\rm sin}\theta }\frac{\partial }{\partial \varphi
}\label{eq3}.
\end{equation}
For a spacetime viewpoint rather than a 3 + 1 split of spacetime,
the set of orthonormal vectors also includes the basis vector for
the time coordinate given by
\begin{equation}
{\bf e}_{\hat 0}=\frac{d}{d\tau }=\frac{1}{\alpha }\frac{\partial
}{\partial t}\label{eq4},
\end{equation}
where $\alpha $ is the lapse function (or redshift factor) defined by
\begin{equation}
\alpha (r)\equiv \frac{d\tau
}{dt}=\left(1-\frac{r^2}{\ell^2}\right)^{1/2}\label{eq5}.
\end{equation}

The gravitational acceleration felt by a FIDO is given by
\cite{thirty five,thirty six,thirty seven,thirty eight}
\begin{equation}
{\bf a}=-\nabla {\rm ln}\alpha
=\frac{1}{\alpha}\frac{r}{\ell^2}\,{\bf e}_{\hat r}\label{eq6},
\end{equation}
while the rate of change of any scalar physical quantity or any
three-dimensional vector or tensor, as measured by a FIDO, is
defined by the convective derivative
\begin{equation}
\frac{D}{D\tau }\equiv \left(\frac{1}{\alpha }\frac{\partial
}{\partial t}+{\bf v}\cdot \nabla \right)\label{eq7},
\end{equation}
$\bf v$ being the velocity of a fluid as measured locally by a FIDO.

\section{Two-Fluid Plasma Equations in 3+1 Formalism}\label{sec3}
We consider a two-component plasma consisting of electrons and
either positrons or ions. In the 3+1 notation, the continuity
equation for each of the fluid species is
\begin{equation}
\frac{\partial }{\partial t}(\gamma _sn_s)+\nabla \cdot (\alpha
\gamma _sn_s{\bf v}_s)=0\label{eq8}.
\end{equation}
where $s$ is 1 for electrons and 2 for positrons (or ions). For a
perfect relativistic fluid of species $s$ in three-dimensions, the
energy density $\epsilon_s$, the momentum density ${\bf S}_s$, and
stress-energy tensor $W^{jk}_ s$ are given by
\begin{equation}
\epsilon _s=\gamma _s^2(\varepsilon _s+P_s{\bf
v}_s^2),\hspace{0.6cm} {\bf S}_s=\gamma _s^2(\varepsilon _s+P_s){\bf
v}_s,\hspace{0.6cm} W_s^{jk}=\gamma _s^2(\varepsilon
_s+P_s)v_s^jv_s^k+P_sg^{jk}.\label{eq9}
\end{equation}
where ${\bf v}_s$ is the fluid velocity, $n_s$ is the number
density, $P_s$ is the pressure, and $\varepsilon _s$ is the total
energy density defined by
\begin{equation}
\varepsilon _s=m_sn_s+P_s/(\gamma _g-1)\label{eq10}.
\end{equation}
The gas constant $\gamma _g$ is $4/3$ for $T\rightarrow \infty $ and
$5/3$ for $T\rightarrow 0$.

Using the conservation of entropy, the equation of state can be
expressed by
\begin{equation}
\frac{D}{D\tau }\left(\frac{P_s}{n_s^{\gamma
_g}}\right)=0\label{eq11},
\end{equation}
where $D/D\tau =(1/\alpha )\partial /\partial t+{\bf v}_s\cdot
\nabla $. The full equation of state for a relativistic fluid, as
measured in the fluid's rest frame, is as follows \cite{thirty
nine,fourty}:
\begin{equation}
\varepsilon
=m_sn_s+m_sn_s\left[\frac{P_s}{m_sn_s}-\frac{\textrm{i}H_2^{(1)^\prime
}(\textrm{i}m_sn_s/P_s)}{\textrm{i}H_2^{(1)}(\textrm{i}m_sn_s/P_s)}\right]\label{eq12},
\end{equation}
where the $H_2^{(1)}(x)$ are Hankel functions.

The quantities of (\ref{eq9}) in the electromagnetic field are
expressed by
\begin{eqnarray}
\epsilon_s&=&\frac{1}{8\pi }({\bf E}^2+{\bf B}^2), \hspace{1cm}{\bf
S}_s=\frac{1}{4\pi }{\bf E}\times {\bf B}\nonumber,\\
W_s^{jk}&=&\frac{1}{8\pi }({\bf E}^2+{\bf B}^2)g^{jk}-\frac{1}{4\pi
}(E^jE^k+B^jB^k)\label{eq13}.
\end{eqnarray}
The equations for the conservation of energy and momentum are
respectively given by \cite{thirty five,thirty six,thirty seven}
\begin{eqnarray}
\frac{1}{\alpha }\frac{\partial }{\partial t}\epsilon _s&
=&-\nabla \cdot {\bf S}_s+2{\bf a}\cdot {\bf S}_s,\label{eq14}\\
\frac{1}{\alpha }\frac{\partial }{\partial t}{\bf S}_s&=&\epsilon
_s{\bf a}-\frac{1}{\alpha }\nabla \cdot (\alpha
{\stackrel{\leftrightarrow }{\bf W}}_s)\label{eq15}.
\end{eqnarray}
When the two-fluid plasma couples to the electromagnetic fields, the
Maxwell's equations take the following 3+1 form:
\begin{eqnarray}
\nabla \cdot {\bf B}&=&0,\label{eq16}\\
\nabla \cdot {\bf E}&=&4\pi \sigma ,\label{eq17}\\
\frac{\partial {\bf B}}{\partial t}&=&-\nabla \times (\alpha {\bf E}),\label{eq18}\\
\frac{\partial {\bf E}}{\partial t}&=&\nabla \times (\alpha {\bf
B})-4\pi \alpha {\bf J}\label{eq19},
\end{eqnarray}
where the charge and current densities are respectively defined by
\begin{equation}
\sigma =\sum_s\gamma _sq_sn_s,\hspace{1.2cm}{\bf J}=\sum_s\gamma
_sq_sn_s{\bf v}_s\label{eq20}.
\end{equation}

Using (\ref{eq10}) and (\ref{eq16}--\ref{eq19}), the energy and
momentum conservation equations (\ref{eq14}) and (\ref{eq15}) can be
rewritten for each species $s$ in the form
\begin{eqnarray}
\frac{1}{\alpha }\frac{\partial }{\partial t}P_s-\frac{1}{\alpha
}\frac{\partial }{\partial t}[\gamma _s^2 (\varepsilon
_s+P_s)]-\nabla \cdot
[\gamma _s^2(\varepsilon _s+P_s){\bf v}_s]\nonumber\\
+\gamma _sq_sn_s{\bf E}\cdot {\bf v}_s+2\gamma _s^2(\varepsilon _s
+P_s){\bf a}\cdot {\bf v}_s=0\label{eq21},
\end{eqnarray}
\begin{eqnarray}
\gamma _s^2(\varepsilon _s+P_s)\left(\frac{1}{\alpha }\frac{\partial
}{\partial t}+{\bf v}_s\cdot \nabla \right) {\bf v}_s+\nabla
P_s-\gamma _sq_sn_s({\bf E} +{\bf v}_s\times {\bf B})\nonumber\\
+{\bf v}_s\left(\gamma _sq_sn_s{\bf E}\cdot {\bf v}_s
+\frac{1}{\alpha }\frac{\partial }{\partial t}P_s\right)+\gamma
_s^2(\varepsilon _s+P_s)[{\bf v}_s({\bf v}_s\cdot {\bf a})-{\bf
a}]=0\label{eq22}.
\end{eqnarray}
Although these equations are valid in a FIDO frame, they reduce for
$\alpha=1$ to the corresponding special relativistic equations
\cite{fourty one} which are valid in a frame in which both fluids
are at rest. The transformation from the FIDO frame to the comoving
(fluid) frame involves a boost velocity, which is simply the
freefall velocity, given by
\begin{equation} v_{\rm ff}=(1-\alpha
^2)^{\frac{1}{2}}\label{eq23}.
\end{equation}
Then the relativistic Lorentz factor $\gamma _{\rm boost}\equiv
(1-v_{\rm ff}^2)^{-1/2}=1/\alpha $.

For a good approximation near the horizon, we write the dS metric in
the Rindler coordinate system as follows:
\begin{equation}
ds^2=-\left(1-\frac{r^2}{\ell^2}\right)dt^2+dx^2+dy^2+dz^2,\label{eq24}
\end{equation}
where
\begin{equation}
x=\ell\left(\theta -\frac{\pi }{2}\right),\hspace{1cm}y= \ell\varphi
,\hspace{1cm}z=2\ell
\left(1-\frac{r^2}{\ell^2}\right)^{1/2}\label{eq25}.
\end{equation}
The standard lapse function in Rindler coordinates becomes $\alpha =
z/2r_h$, where $r_h=\ell$ is the location of the cosmological event
horizon.

\section{Radial Wave Propagation in One-Dimension}\label{sec4}
We consider one-dimensional wave propagation in the radial $z$
direction and introduce the complex variables
\begin{eqnarray}
v_{sz}(z,t)=u_s(z,t),\hspace{.3cm}
v_s(z,t)=v_{sx}(z,t)+\textrm{i}v_{sy}(z,t),\nonumber\\
B(z,t)=B_x(z,t)+\textrm{i}B_y(z,t),\hspace{.3cm}E(z,t)=E_x(z,t)+\textrm{i}E_y(z,t)\label{eq26}.
\end{eqnarray}
Then
\begin{eqnarray}
v_{sx}B_y-v_{sy}B_x&=&\frac{\textrm{i}}{2}(v_sB^\ast -v_s^\ast B),\nonumber\\
v_{sx}E_y-v_{sy}E_x&=&\frac{\textrm{i}}{2}(v_sE^\ast -v_s^\ast
E)\label{eq27},
\end{eqnarray}
where the $\ast $ denotes the complex conjugate. The continuity
equation (\ref{eq8}) takes the form
\begin{equation}
\frac{\partial}{\partial t}(\gamma _sn_s)+\frac{\partial }{\partial
z}(\alpha \gamma _sn_su_s)=0,\label{eq28}
\end{equation}
while Poisson's equation (\ref{eq17}) becomes
\begin{equation}
\frac{\partial E_z}{\partial z}=4\pi (q_1n_1\gamma _1+q_2n_2\gamma
_2).\label{eq29}
\end{equation}

The ${\bf e}_{\hat x}$ and ${\bf e}_{\hat y}$ components of
(\ref{eq18}) and (\ref{eq19}) give
\begin{eqnarray}
\frac{1}{\alpha }\frac{\partial B}{\partial t}
&=&-\textrm{i}\left(\frac{\partial }{\partial z}-a\right)E,\label{eq30}\\
\textrm{i}\frac{\partial E}{\partial t} &=&-\alpha
\left(\frac{\partial }{\partial z}-a\right) B-\textrm{i}4\pi e\alpha
(\gamma _2n_2v_2-\gamma _1n_1v_1)\label{eq31}.
\end{eqnarray}
Differentiating equation (\ref{eq31}) with respect to $t$ and using
(\ref{eq30}), we obtain
\begin{equation}
\left(\alpha ^2\frac{\partial ^2}{\partial z^2}+\frac{3\alpha}{2r_h
} \frac{\partial }{\partial z}-\frac{\partial ^2 }{\partial
t^2}+\frac{1}{4r_h^2}\right)E=4\pi e\alpha \frac{\partial }{\partial
t}(n_2\gamma _2v_2-n_1\gamma _1v_1)\label{eq32}.
\end{equation}
The transverse component of the momentum conservation equation is
obtained from the ${\bf e}_{\hat x}$ and ${\bf e}_{\hat y}$
components of (\ref{eq22}) as follows:
\begin{equation}
\rho _s\frac{Dv_s}{D\tau }=q_sn_s\gamma _s(E-\textrm{i}v_sB_z
+\textrm{i}u_sB)-u_sv_s\rho _sa -v_s\left(q_sn_s\gamma _s{\bf
E}\cdot {\bf v}_s +\frac{1}{\alpha }\frac{\partial P_s}{\partial
t}\right),\label{eq33}
\end{equation}
where
\[
{\bf E}\cdot {\bf v}_s=\frac{1}{2}(Ev_s^\ast +E^\ast v_s)+E_zu_s
\]
and $\rho _s$ is the total energy density defined by
\begin{equation} \rho _s=\gamma _s^2(\varepsilon _s+P_s)=\gamma
_s^2(m_sn_s+\Gamma _gP_s)\label{eq34}
\end{equation}
with $\Gamma _g=\gamma _g/(\gamma _g-1)$.

\section{Linearization}\label{sec5}
We use perturbation method to linearize the equations derived in the
preceding section by introducing the quantities
\begin{eqnarray}
u_s(z,t)&=&u_{os}(z)+\delta u_s(z,t),\hspace{.5cm}
v_s(z,t)=\delta v_s(z,t),\nonumber\\
n_s(z,t)&=&n_{os}(z)+\delta n_s(z,t),\hspace{.5cm}
P_s(z,t)=P_{os}(z)+\delta P_s(z,t),\nonumber\\
\rho _s(z,t)&=&\rho _{os}(z)+\delta \rho _s(z,t),\hspace{.5cm}
{\bf E}(z,t)=\delta {\bf E}(z,t),\nonumber\\
{\bf B}_z(z,t)&=&{\bf B}_o(z)+\delta {\bf B}_z(z,t),\hspace{.5cm}
{\bf B}(z,t)=\delta {\bf B}(z,t)\label{eq35},
\end{eqnarray}
where magnetic field is chosen to lie along the radial ${\bf
e}_{\hat z}$ direction. The relativistic Lorentz factor is also
linearized such that
\begin{equation}
\gamma _s=\gamma _{os}+\delta \gamma _s,\qquad\mbox{where}\quad
\gamma _{os}=\left(1-{\bf
u}_{os}^2\right)^{-\frac{1}{2}},\quad\delta \gamma _s=\gamma
_{os}^3{\bf u}_{os}\cdot \delta {\bf u}_s\label{eq36}.
\end{equation}
Near the horizon the unperturbed radial velocity for each species as
measured by a FIDO along ${\bf e}_{\hat z}$ is assumed to be the
freefall velocity so that
\begin{equation}
u_{os}(z)=v_{\textrm{ff}}(z)=[1-\alpha
^2(z)]^{\frac{1}{2}}\label{eq37}.
\end{equation}
It follows, from the continuity equation (\ref{eq28}), that
\[
r^2\alpha \gamma _{os}n_{os}u_{os}=\mbox{const.}=r_h^2\alpha
_h\gamma _hn_hu_h,
\]
where the values with a subscript $h$ are the limiting values at the
horizon. The freefall velocity at the horizon becomes unity so that
$u_h=1$. Since $u_{os}=v_{\textrm{ff}}$, $\gamma _{os}=1/\alpha $;
hence, $\alpha \gamma _{os}=\alpha _h\gamma _h=1$. Also, because
$v_{\textrm{ff}}=r/r_h$, the number density for each species can be
written as follows:
\begin{equation}
n_{os}(z)=n_{hs}v_{\textrm{ff}}^{-3}\label{eq38}.
\end{equation}
The equation of state (\ref{eq11}) and (\ref{eq38}) lead to write
the unperturbed pressure, in terms of the freefall velocity, as
follows:
\begin{equation}
P_{os}(z)=P_{hs}v_{\textrm{ff}}^{-3\gamma _g}\label{eq39}.
\end{equation}
Since $P_{os}=k_Bn_{os}T_{os}$, then with $k_B=1$, the temperature
profile is
\begin{equation}
T_{os}=T_{hs}v_{\rm ff}^{-3(\gamma _g-1)}(z)\label{eq40}.
\end{equation}
The unperturbed magnetic field is purely in the radial direction and
it does not experience effects of spatial curvature. From the flux
conservation $\nabla \cdot {\bf B}_o=0$ it follows that
\[
r^2B_o(r)=\mbox{const.}
\]
One can obtain from this the unperturbed magnetic field, in terms of
the freefall velocity, as follows:
\begin{equation}
B_o(z)=B_hv_{\textrm{ff}}^{-2},\label{eq41}
\end{equation}
where $v_{\textrm{ff}}=[1-\alpha ^2(z)]^{1/2}$. Since
\begin{equation}
\frac{dv_{\textrm{ff}}}{dz}=-\frac{\alpha }{2r_h}\frac{1}{v_{\rm
ff}},\label{eq42}
\end{equation}
we have
\begin{eqnarray}
\frac{du_{os}}{dz}&=&-\frac{\alpha }{2r_h}\frac{1}{v_{\textrm{ff}}
},\qquad \frac{dB_o}{dz}=\frac{\alpha }{r_h}\frac{B_o}{v_{\textrm{ff}}^2},\nonumber\\
\frac{dn_{os}}{dz}&=&\frac{3\alpha }{2r_h}n_{os},\qquad
\frac{dP_{os}}{dz}=\frac{3\alpha }{2r_h}\frac{\gamma
_gP_{os}}{v_{\rm ff}^2}\label{eq43}.
\end{eqnarray}

When the linearized variables from (\ref{eq35}) and (\ref{eq36}) are
substituted into the continuity equation and products of
perturbation terms are neglected, the result gives
\begin{eqnarray}
\gamma_{os}\left(\frac{\partial}{\partial
t}+u_{os}\alpha\frac{\partial}{\partial
z}+\frac{u_{os}}{2r_h}+\gamma_{os}^2\alpha\frac{du_{os}}{dz}\right)\delta
n_s+\left(\alpha\frac{\partial}{\partial
z}+\frac{1}{2r_h}\right)(n_{os}\gamma_{os}u_{os})\nonumber\\
+n_{os}\gamma_{os}^3\left[u_{os}\frac{\partial}{\partial
t}+\alpha\frac{\partial}{\partial z}+\frac{1}{2r_h}+\alpha
\left(\frac{1}{n_{os}}\frac{dn_{os}}{dz}+3\gamma_{os}^2u_{os}
\frac{du_{os}}{dz}\right)\right]\delta u_s=0.\label{eq44}
\end{eqnarray}
In the similar way, we obtain from the conservation of entropy,
(\ref{eq11}),
\begin{equation}
\delta P_s=\frac{\gamma_gP_{os}}{n_{os}}\delta n_s\label{eq45},
\end{equation}
and from the total energy density, (\ref{eq34}),
\begin{equation}
\delta \rho_s=\frac{\rho_{os}}{n_{os}}\left(1
+\frac{\gamma_{os}^2\gamma_gP_{os}}{\rho_{os}}\right)\delta
n_s+2u_{os}\gamma_{os}^2\rho_{os}\delta u_s\label{eq46},
\end{equation}
where $\rho_{os}=\gamma_{os}^2(m_sn_{os}+\Gamma_gP_{os})$.
Linearizing the transverse part of the momentum conservation
equation, differentiating it with respect to $t$ and then
substituting from (\ref{eq30}), we obtain
\begin{eqnarray}
\left(\alpha u_{os}\frac{\partial }{\partial z} +\frac{\partial
}{\partial t}-\frac{u_{os}}{2r_h}+\frac{\textrm{i}\alpha q_s\gamma
_{os}n_{os}B_o}{\rho _{os}}\right)\frac{\partial \delta v_s}{\partial t}\nonumber\\
-\frac{\alpha q_s\gamma _{os}n_{os}}{ \rho _{os}}\left(\alpha
u_{os}\frac{\partial }{
\partial z}+\frac{\partial }{\partial t}-\frac{u_{os}}{2r_h}\right)\delta E=0\label{eq47}.
\end{eqnarray}
When linearized, Poisson's equation (\ref{eq29}) and (\ref{eq32})
respectively give
\begin{eqnarray}
\frac{\partial\delta E_z}{\partial z}&=&4\pi
e(n_{o2}\gamma_{o2}-n_{o1}\gamma_{o1})
+4\pi e(\gamma_{o2}\delta n_2-\gamma_{o1}\delta n_1)\nonumber\\
&&+4\pi e(n_{o2}u_{o2}\gamma_{o2}^3\delta
u_2-n_{o1}u_{o1}\gamma_{o1}^3\delta u_1)\label{eq48},
\end{eqnarray}
\begin{equation}
\left(\alpha ^2\frac{\partial ^2}{\partial z^2}+\frac{3\alpha}{2r_h
} \frac{\partial }{\partial z}-\frac{\partial ^2 }{\partial
t^2}+\frac{1}{4r_h^2}\right)\delta E=4\pi e\alpha
\left(n_{o2}\gamma_{o2}\frac{\partial \delta v_2}{\partial
t}-n_{o1}\gamma_{o1}\frac{\partial \delta v_1}{\partial
t}\right)\label{eq49}.
\end{equation}

\section{Dispersion Relation}\label{sec6}
Our consideration effects on a local scale for which the distance
from the horizon does not vary significantly. We use a local (or
mean-field) approximation for the lapse function and hence for the
equilibrium fields and fluid quantities. If the plasma is situated
relatively close to the horizon, $\alpha ^2\ll 1$, then a relatively
small change in distance $z$ will make a significant difference to
the magnitude of $\alpha $. Thus it is important to choose a
sufficiently small range in $z$ so that $\alpha $ does not vary
much.

We consider thin layers in the ${\bf e}_{\hat z}$ direction, each
layer with its own $\alpha_o$, where $\alpha_o$ is some mean value
of $\alpha$ within a particular layer. Then a more complete picture
can be built up by considering a large number of layers within a
chosen range of $\alpha_o$ values.

The local approximation imposes the restriction that the wavelength
must be smaller in magnitude than the scale of the gradient of the
lapse function $\alpha$, i.e., $\lambda<(\partial \alpha /\partial
z)^{-1}=2\ell$, or equivalently, $k>(\pi/\ell)$.

One of the disadvantages of the hydrodynamical approach is that it
is essentially a bulk, fluid approach and therefore the microscopic
behavior of the two-fluid plasma is treated in a somewhat
approximate manner via the equation of state. It means that the
results are really only strictly valid in the long wavelength limit.
However, the restriction, imposed by the local approximation, on the
wavelength is not too severe and permits the consideration of
intermediate to long wavelengths so that the small $k$ limit is
still valid.

In the local approximation for $\alpha$, $\alpha\simeq\alpha_o$ is
valid within a particular layer. Hence, the unperturbed fields and
fluid quantities and their derivatives, which are functions of
$\alpha$, take on their corresponding \lq \lq  mean-field\rq \rq
values for a given $\alpha_o$. Then the coefficients in
(\ref{eq44}), (\ref{eq47}) and (\ref{eq48}) are constants within
each layer with respect to $\alpha$ (and therefore $z$ as well).
Hence, it is possible to Fourier transform the equations with
respect to $z$, assuming plane-wave-type solutions for the
perturbations of the form $\sim e^{(kz-\omega t)}$ for each
$\alpha_o$ layer.

When Fourier transformed, (\ref{eq47}) and (\ref{eq49}) turn out to
be
\begin{equation}
\delta E=\frac{\textrm{i}4\pi
e\alpha_o\omega(n_{o2}\gamma_{o2}\delta v_2-n_{o1}\gamma_{o1}\delta
v_1)}{\alpha_ok(\alpha_ok-\textrm{i}3/2r_h)-\omega^2-1/(2r_h)^2},\label{eq50}
\end{equation}
\begin{equation}
\omega \left(\alpha _oku_{os}-\omega +\frac{\textrm{i}u_{os}}{2r_h}
+\frac{\alpha _oq_s\gamma _{os}n_{os}B_o}{\rho _{os}}\right)\delta
v_s-\textrm{i}\alpha _o\frac{q_s\gamma _{os}n_{os}}{\rho
_{os}}\left(\alpha _oku_{os}-\omega
-\frac{\textrm{i}u_{os}}{2r_h}\right)\delta E=0\label{eq51}.
\end{equation}
The dispersion relation for the transverse electromagnetic wave
modes may be written as
\begin{eqnarray}
\left[K_\pm \left(K_\pm \pm \frac{\textrm{i}}{2r_h}\right) -\omega
^2+\frac{1}{(2r_h)^2}\right] =\alpha _o^2\left\{\frac{\omega
_{p1}^2(\omega -u_{o1}K_\pm )}{(u_{o1}K_\mp -\omega -\alpha _o\omega
_{c1})}+\frac{\omega _{p2}^2(\omega -u_{o2}K_\pm )}{(u_{o2}K_\mp
-\omega +\alpha _o\omega _{c2})}\right\}\label{eq52}
\end{eqnarray}
for either the electron-positron or electron-ion plasma. Here,
$K_\pm =\alpha _ok\pm \textrm{i}/2r_h$,
$\omega_{cs}=e\gamma_{os}n_{os}B_o/\rho_{os}$, and
$\omega_{ps}=\sqrt{4\pi e^2\gamma_{os}^2n_{os}^2/\rho_{os}}$. The
cyclotron frequency $\omega_{cs}$, as well as the plasma frequency
$\omega_{ps}$, is frame independent. Although the fluid quantities
are measured in the fluid frame, the field $B_o$ is measured in the
FIDO frame. Hence, the factors of $\gamma_{os}$ do not cancel out
explicitly. The transformation $B_o\rightarrow\gamma_{os}B_o$ boosts
the fluid frame for either fluid and thereby cancels the
$\gamma_{os}$ factors. The $+$ and $-$ denote the left $L$ and right
$R$ modes, respectively. The complex conjugate of the dispersion
relation for the $R$ mode gives the dispersion relation for the $L$
mode. In the special relativistic case, the two modes have the same
dispersion relation.

\section{Numerical Solution Modes}\label{sec7}
The dispersion relations (\ref{eq52}) are complicated enough even in
the simplest cases for the electron-positron plasma where both
species are assumed to have the same equilibrium parameters, and an
analytical solution is cumbersome and unprofitable. We therefore
solve numerically the dispersion relation in order to determine all
the physically meaningful modes for the transverse waves. We put the
equations in the form of a matrix equation as follows:
\begin{equation}
(A-kI)X=0\label{eq53},
\end{equation}
where the eigenvalue is chosen to be the wave number $k$, the
eigenvector $X$ is given by the relevant set of perturbations, and
$I$ is the identity matrix.

In order to write the perturbation equations in an appropriate form,
we introduce the following set of dimensionless variables:
\begin{eqnarray}
\tilde \omega =\frac{\omega }{\alpha _o\omega _\ast },\quad
\tilde k=\frac{kc}{\omega _\ast },\quad k_h=\frac{1}{2r_h\omega _\ast },\nonumber\\
\delta \tilde u_s=\frac{\delta u_s}{u_{os}},\quad \tilde v_s
=\frac{\delta v_s}{u_{os}},\quad \delta \tilde n_s=\frac{\delta n_s}{n_{os}},\nonumber\\
\delta \tilde B=\frac{\delta B}{B_o},\quad \tilde E=\frac{\delta
E}{B_o},\quad \delta \tilde E_z=\frac{\delta E_z}{B_o}.\label{eq54}
\end{eqnarray}
For an electron-positron plasma, $\omega _{p1}=\omega _{p2}$ and
$\omega _{c1}=\omega _{c2}$; so, $\omega _\ast $ is defined as
\begin{equation}
\omega _\ast =\left\{\begin{array}{rl}&\omega _c,\quad\mbox{Alfv\'en modes},\\
&\\
&(2\omega _p^2+\omega _c^2)^{\frac{1}{2}},\quad \mbox{high frequency
modes},\end{array}\right.\label{eq55}
\end{equation}
where $\omega _p=\sqrt{\omega _{p1}\omega _{p2}}$ and $\omega
_c=\sqrt{\omega _{c1}\omega _{c2}}$. However, for the case of an
electron-ion plasma, the plasma frequency and the cyclotron
frequency are different for each fluid; so, the choice of $\omega
_\ast $ is a more complicated matter. We assume, for simplicity,
that

\begin{equation}
\omega _\ast =\left\{\begin{array}{rl}&\frac{1
}{\sqrt{2}}(\omega _{c1}^2+\omega _{c2}^2)^{\frac{1}{2}},\quad \mbox{Alfv\'en modes},\\
&\\
&(\omega _{\ast 1}^2+\omega _{\ast 2}^2)^{\frac{1}{2}},\qquad
\mbox{high frequency modes},\end{array}\right.\label{eq56}
\end{equation}
where $\omega _{\ast s}^2=(2\omega _{ps}^2+\omega _{cs}^2)$.

The dimensionless eigenvector for the transverse set of equations is
\begin{equation}
\tilde X_{\rm transverse}=\left[\begin{array}{c}\delta \tilde
v_1\\\delta \tilde v_2\\\delta \tilde B\\\delta \tilde
E\end{array}\right]\label{eq57}.
\end{equation}
When linearized and Fourier transformed, equations (\ref{eq30}) and
(\ref{eq31}) turn out to be
\begin{equation}
\left(k-\frac{\textrm{i}}{2r_h\alpha_o}\right)\delta
E+\frac{\textrm{i}\omega}{\alpha_o}\delta B=0\label{eq58},
\end{equation}
\begin{equation}
\frac{\textrm{i}\omega}{\alpha_o}\delta
E=\left(k-\frac{\textrm{i}}{2r_h\alpha_o}\right)\delta B+4\pi
e(\gamma_{o2} n_{o2}\delta v_2-\gamma_{o1} n_{o1}\delta
v_1)\label{eq59}.
\end{equation}
Using (\ref{eq54}), we write (\ref{eq51}), (\ref{eq58}), and
(\ref{eq59}) in the dimensionless form:
\begin{eqnarray}
\tilde k\delta \tilde v_s&=&\left(\frac{\tilde \omega }{
u_{os}}-\left(\frac{q_s}{e}\right)\frac{\omega _{cs}}{ u_{os}\omega
_\ast }-\frac{\textrm{i}k_h}{ \alpha _o}\right)\delta \tilde v_s
+\left(\frac{q_s}{e}\right)\frac{\omega _{cs}}{ u_{os}\omega _\ast
}\delta \tilde B-\textrm{i}\left(\frac{q_s}{
e}\right)\frac{\omega _{cs}}{u_{os}\omega _\ast }\delta \tilde E,\label{eq60}\\
\tilde k\delta \tilde E&=&-\textrm{i}\tilde \omega \delta \tilde B
+\frac{\textrm{i}k_h}{\alpha _o}\delta \tilde E,\label{eq61}\\
\tilde k\delta \tilde B&=&u_{o1}\frac{\omega _{p1}^2}{ \omega
_{c1}\omega _\ast }\delta \tilde v_1-u_{o2}\frac{ \omega
_{p2}^2}{\omega _{c2}\omega _\ast }\delta \tilde v_2+\frac{
\textrm{i}k_h}{\alpha _o}\delta \tilde B+\textrm{ i}\tilde \omega
\delta \tilde E \label{eq62}.
\end{eqnarray}
These are the equations in the required form to be used as input to
(\ref{eq53}).

\section{Results}\label{sec8}
We carried out the numerical analysis using the well known MATLAB.
We have considered both the electron-positron plasma and the
electron-ion plasma. The limiting horizon values for the
electron-positron plasma are taken to be
\begin{equation}
n_{hs}=10^{18}\textrm{cm}^{-3},\quad T_{hs }=10^{10}\textrm{K},
\quad B_{h}=3\times 10^6\textrm{G},\quad\mbox{and}\quad\gamma
_g=\frac{4}{3}\label{eq63}.
\end{equation}
For the electron-ion plasma, the ions are essentially non
relativistic, and the limiting horizon values are chosen to be
\begin{equation}
n_{h1}=10^{18}\textrm{cm}^{-3},\quad T_{h1}=10^{10}\textrm{K};\quad
n_{h2}=10^{15}\textrm{cm}^{-3},\quad
T_{h2}=10^{12}\textrm{K}\label{eq64}.
\end{equation}
The equilibrium magnetic field has the same value as it has for the
electron-positron case. The gas constant is $\gamma _g=4/3$.

\subsection{Alfv\'en Modes}\label{subsec8.1}
\subsubsection{Electron-Positron Plasma}\label{subsubsec8.1.1}
For the ultrarelativistic electron-positron plasma in the special
relativistic case, only one purely real Alfv\'en mode exists
\cite{fourty one}, while for the Schwarzschild case there are two
Alfv\'en modes \cite{thirty three}. In our analysis we find three
Alfv\'en modes, as shown in Fig. 1 and Fig. 2., for the
electron-positron plasma. The Alfv\'en modes in de Sitter space are
interesting in that, there exists three Alfv\'en modes for the
electron-positron plasma compared with four modes for the
electron-ion plasma. These three modes for the electron-positron
plasma coalesce into a single mode on taking the special
relativistic limit, giving the result of ref. \cite{fourty one}.
Since we have used the convention $e^{\textrm{i}kz}
=e^{\textrm{i}[\textrm{Re}(k)+\textrm{iIm}(k)]}$, the damping
corresponds to $\textrm{Im}(\tilde{k})>0$ and  growth to
$\textrm{Im}(\tilde{k})<0$.

\subsubsection{Electron-Ion Plasma}\label{subsubsec8.1.2}
In this case four modes are found, two of which are growth and the
other two are damped. The modes shown in Fig. 3 and Fig. 4 are
damped and the remaining two modes shown in Fig. 5 and Fig. 6 are
growth. The first two modes are equivalent to the modes shown in
Fig. 1 and the other two are equivalent to the modes shown in Fig. 2
for electron-positron plasma.

The differences in the magnitudes of the $\omega_{c1}$ and
$\omega_{c2}$ for the first two modes apparently lead to take the
frequencies from their negative (and therefore unphysical) values
for the electron-positron case to positive physical values for the
electron-ion case. These changes are thus because of the difference
in mass and density factors as between the positrons and ions.

These four modes for electron-ion plasma are equivalent to those of
the Schwarzschild case \cite{thirty three}. It is evident that the
growth and damping rates are independent of the frequency, but
depended only on the value of $\alpha_o$.

\subsection{High Frequency Modes}\label{subsec8.2}
\subsubsection{Electron-Positron Plasma}\label{subsubsec8.2.1}
In this case three high frequency electromagnetic modes are found
for the electron-positron plasma, as shown in Figs. 7--9. High
frequency modes in the horizon of dS space for each fluid are
interesting in that all the modes are both damping and growth modes.

The two modes, shown in Figs. 7 and 8, are similar and both modes
are damping and growth very near to the horizon. These two modes are
growth for most of the frequency domain but shows damped for lower
frequencies as $\omega \rightarrow .04$ and $\alpha_o \rightarrow
0$. Thus at a distance from the horizon corresponding to $ \alpha_o
\rightarrow  0.2$ it appears that energy is no longer fed into wave
mode by the gravitational field but begins to be drained from the
waves. The third mode shown in Fig. 9 is also growth and damping
mode. This mode is damped for most of the frequency domain but shows
growth for lower frequencies as $\omega\rightarrow .04$ and
$\alpha_o\rightarrow  0$. These three modes are equivalent to the
three modes of ref. \cite{thirty three} for this case. Also these
three modes coalesce with a single modes in the special relativistic
case \cite{fourty one} as $\alpha_o\rightarrow1$.

\subsubsection{Electron-Ion Plasma}\label{subsubsec8.2.2}
Similar as for the electron-positron plasma, the electron-ion plasma
has three high frequency modes. Two of these, shown in Fig. 10 and
Fig. 11, shows damping and growth very near to the horizon for lower
frequencies, and for upper frequencies they are growth modes. These
two modes are growth for higher frequency but shows damped for lower
frequencies as $\omega \rightarrow .02$ and $\alpha_o\rightarrow 0$.
The third mode, shown in Fig. 12, is growth for lower frequencies as
$\omega \rightarrow .02$ and $\alpha_o \rightarrow 0$ and shows
damped for higher frequencies. These three modes are similar to the
modes for high frequency electron-ion plasma in the Schwarzschild
case \cite{thirty three}.

\section{Concluding Remarks}\label{sec9}
The main concern of this study has been exclusively the
investigation, within the local approximation, of Alfv\'en and high
frequency transverse electromagnetic waves in a two-plasma in the
purely de Sitter space. We derive the dispersion relations for the
Alfv´en and high frequency electromagnetic waves by using a local
approximation and give their numerical solutions. In the limit
$\ell\rightarrow 0$ our results reduce to that in special relativity
as obtained by Sakai and Kawata \cite{fourty one} (i.e., only one
purely real mode for Alfv\'en and high frequency electromagnetic
waves). In contrast to the work of Sakai and Kawata \cite{fourty
one}, new modes (damped or growth) arise for the Alfv´en and high
frequency electromagnetic waves in the pure dS space. In our work
all the modes for Alfv\'en waves are either damped or growing, but
for high frequency electromagnetic waves all the modes are both
damped and growing. This is because of the singularity of the de
Sitter space. For the electron-positron plasma, the damping and
growth rates are similar with the electron-ion plasma but different
by several orders of magnitude, compared with the real components of
the wave number. For both the fluid components the damping and
growth rates are obviously frequency independent, but are dependent
on the radial distance from the horizon as denoted by the mean value
of the lapse function $\alpha _o$. This is of course not for the
case of the high frequency waves. In that case the rate of damping
or growth is dependent on both frequency and radial distance from
the horizon. Damped modes demonstrate, at least in this
approximation, that energy is being drained from some of the waves
by the gravitational field. The majority of the modes are growth
rates and that indicate that the gravitational field is feeding
energy into the waves.

In the light of recent astronomical observations, it has been
suggested that our universe will asymptotically approach a de Sitter
space \cite{one}. Hence, aspects of the de Sitter space might be of
interest in a broader context. Our study of plasmas in the de Sitter
space is thus well motivated.

\vspace{1.0cm}

\noindent
{\large\bf Acknowledgement}\\
One of the authors (MHA) thanks the SIDA as well as the Abdus Salam
International Centre for Theoretical Physics (ICTP), Trieste, Italy,
for supporting with an Associate position of the Centre.

\newpage

\begin{figure}[h]\label{1}
\begin{center}
\includegraphics[scale=0.4]{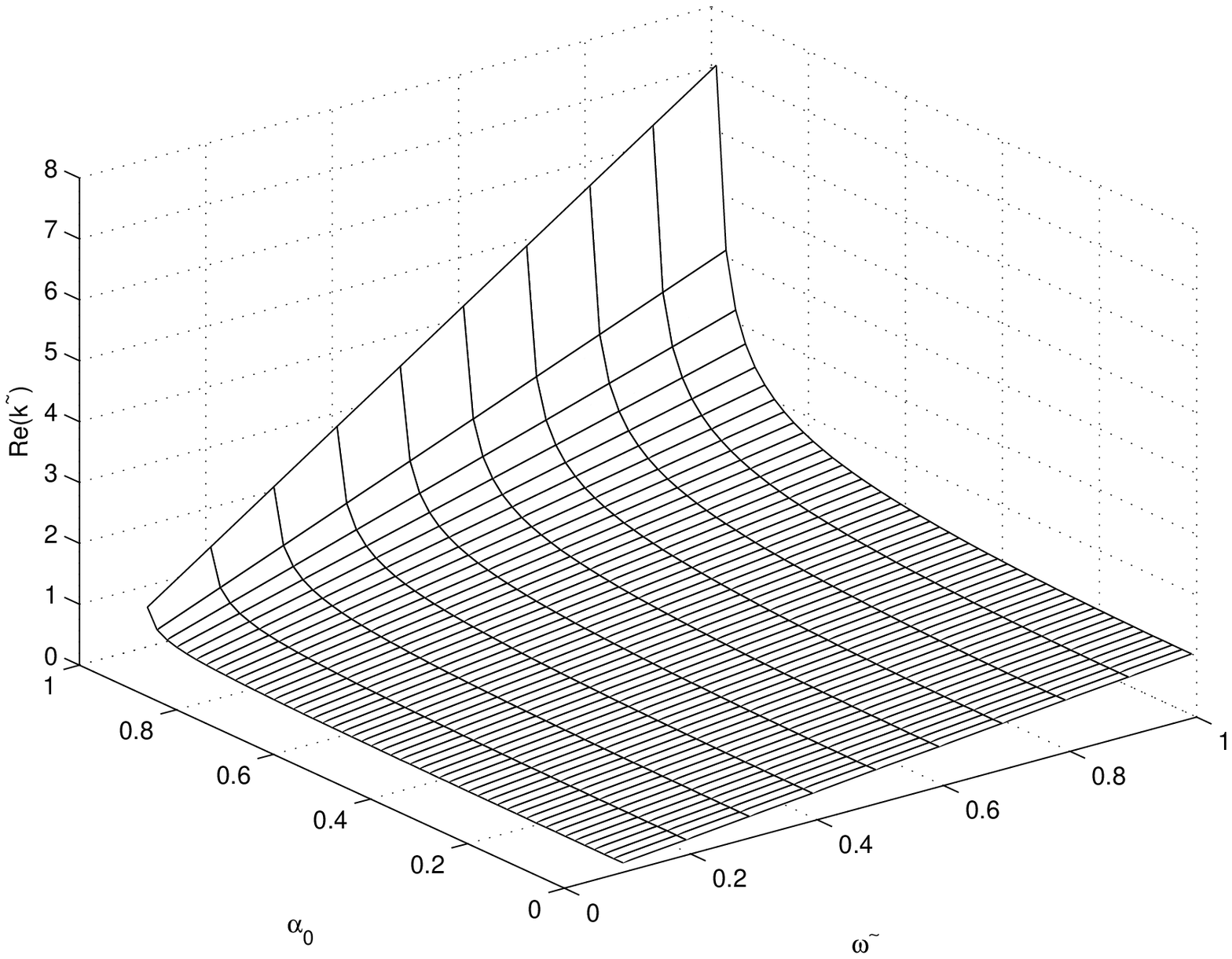}
\end{center}
\begin{center}
\includegraphics[scale=0.4]{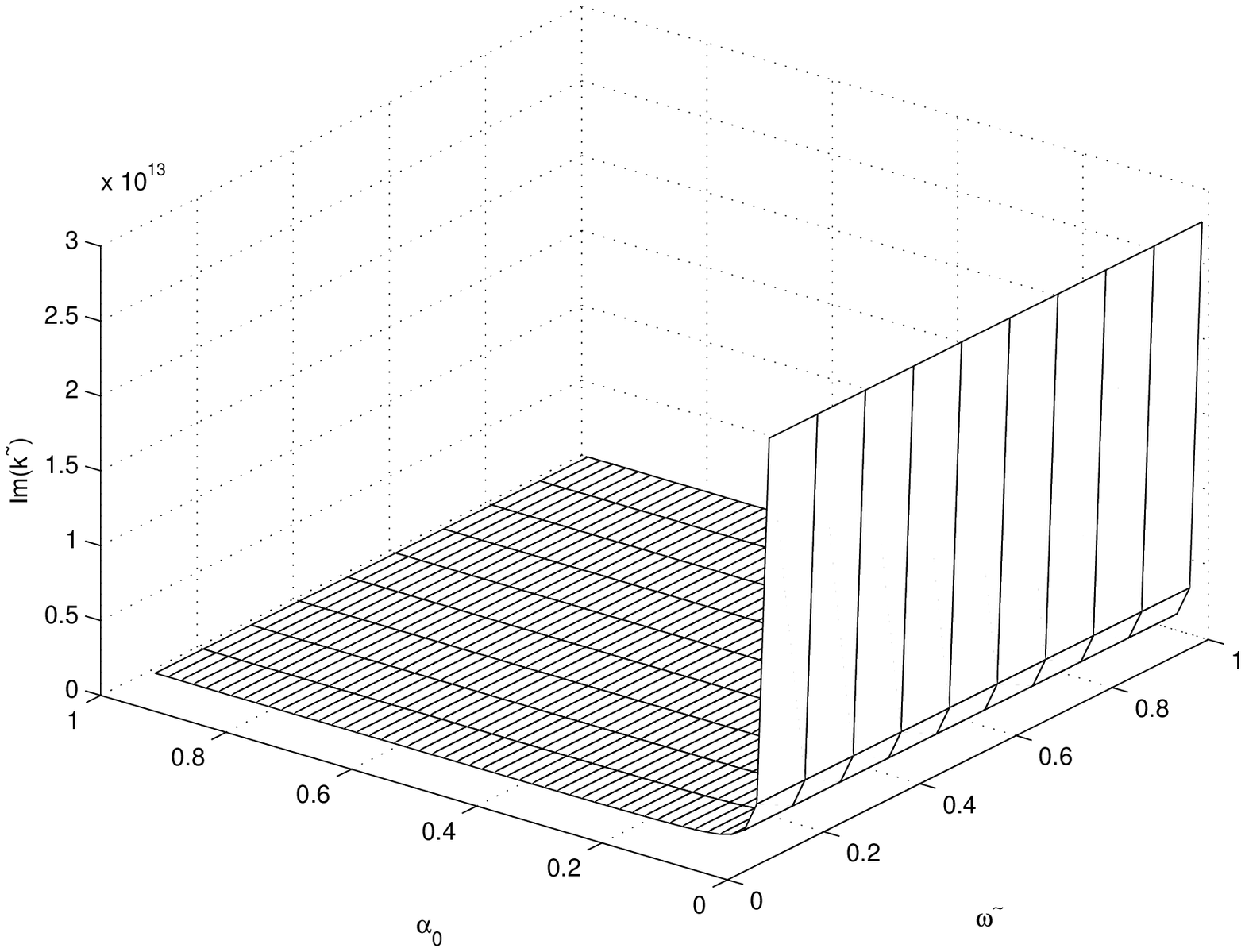}
\end{center}
\begin{center}
\includegraphics[scale=0.4]{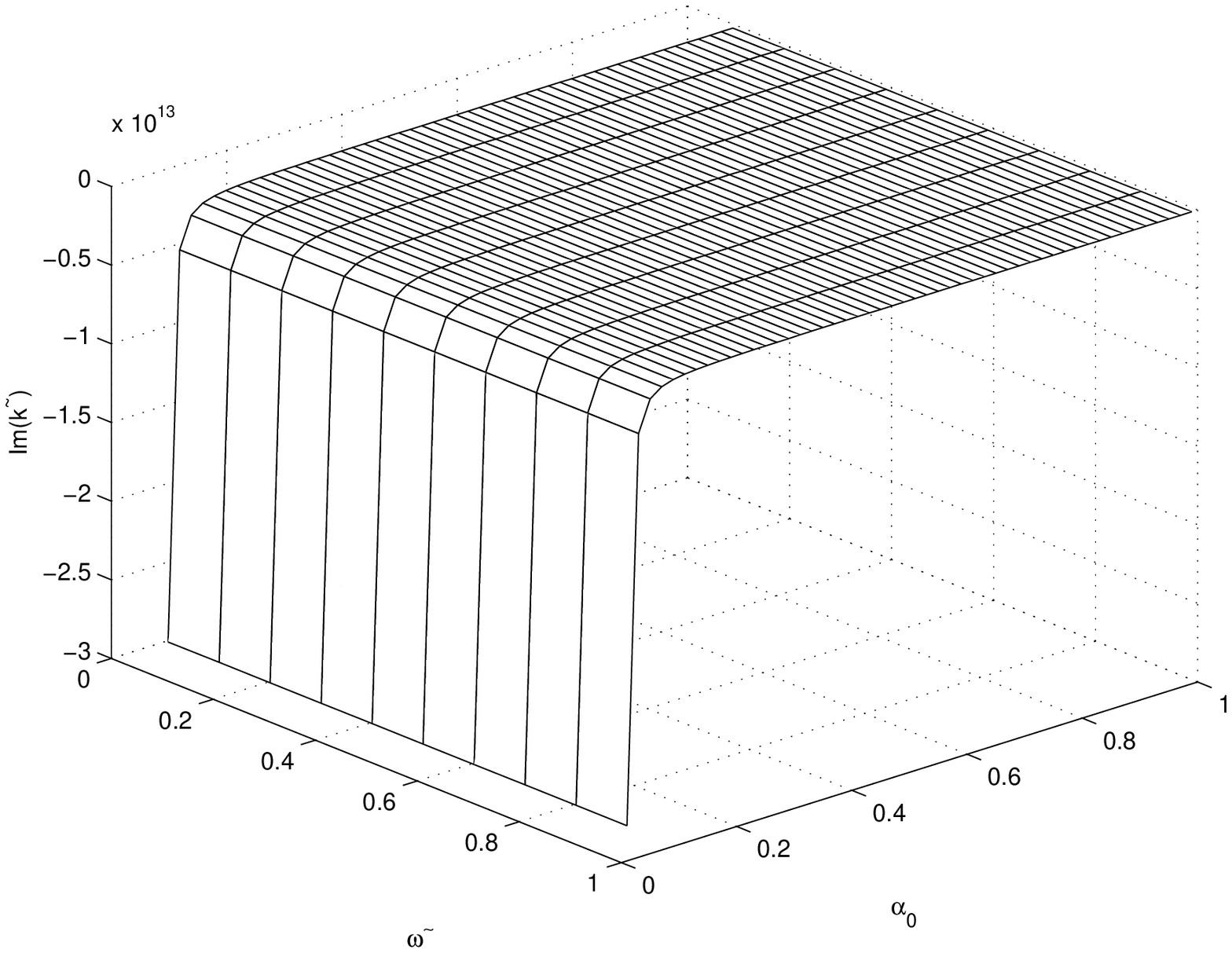}
\end{center}
\caption{\it Top: Real part of Alfv\'en mode for the
electron-positron plasma. Middle: Imaginary part of Alfv\'en damped
mode. Bottom: Imaginary part of Alfv\'en growth mode.}
\end{figure}

\begin{figure}[h]\label{2}
\begin{center}
\includegraphics[scale=.4]{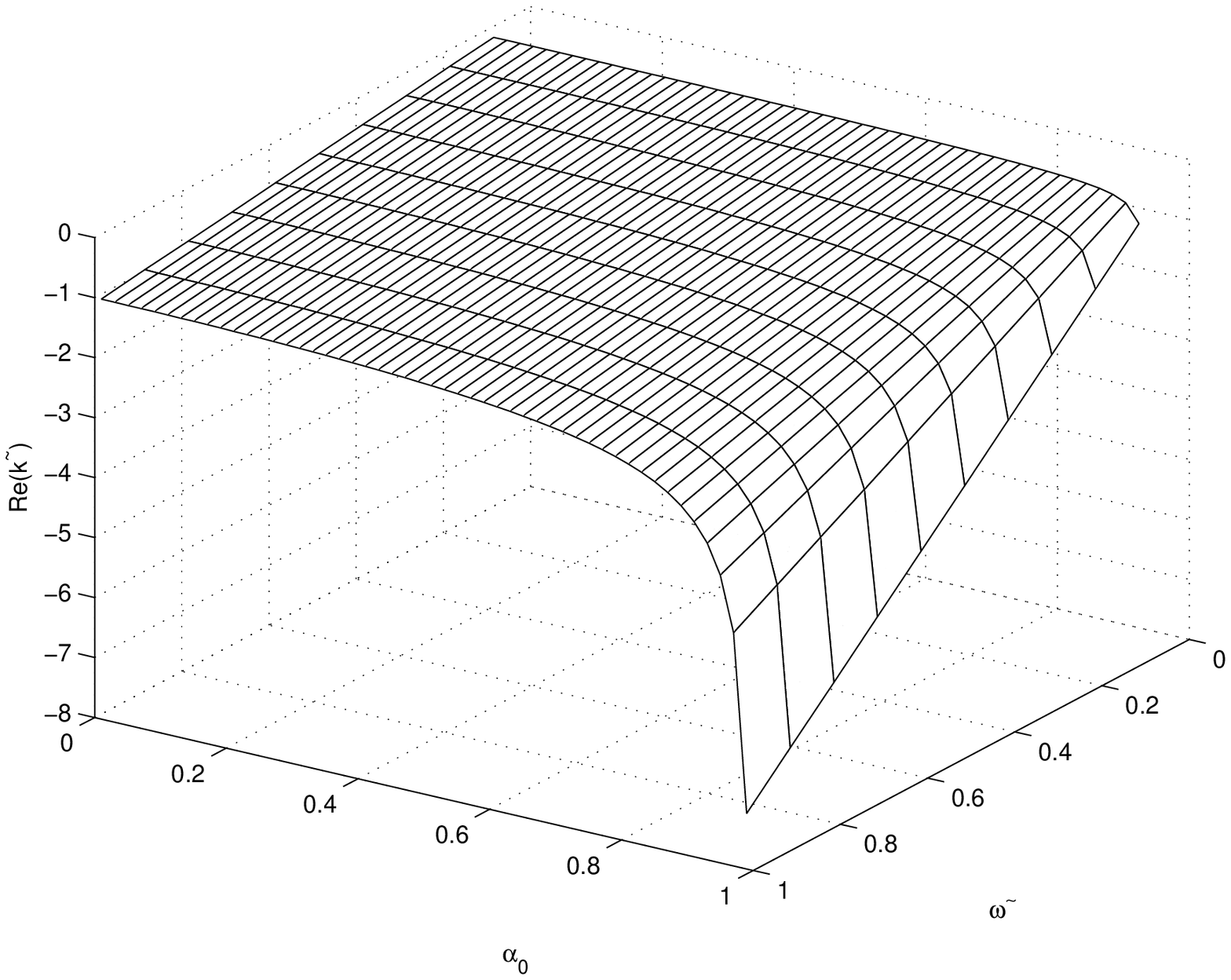}
\includegraphics[scale=.4]{dtalfe_p_imag2.eps}
\end{center}
\caption{\it Left: Real part of Alfv\'en mode for the
electron-positron plasma. Right: Imaginary part of Alfv\'en growth
mode.}
\end{figure}

\begin{figure}[h]\label{3}
\begin{center}
\includegraphics[scale=0.4]{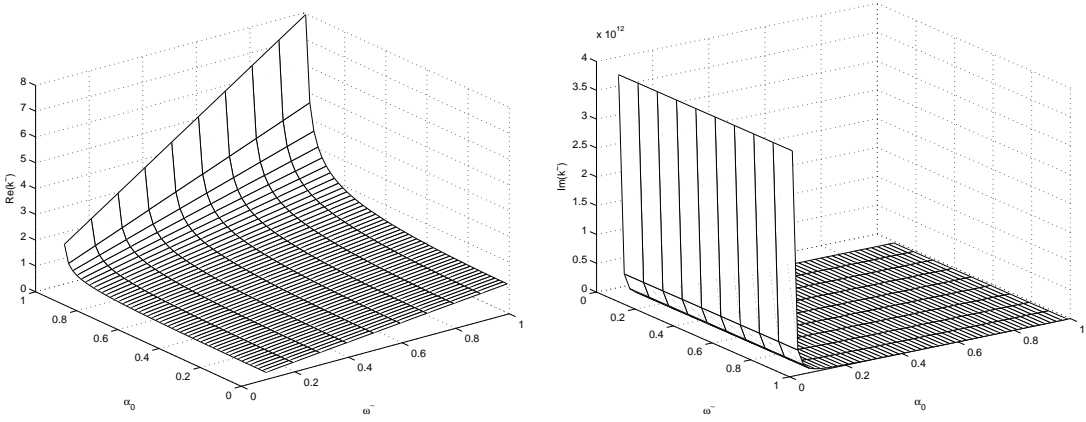}
\includegraphics[scale=0.4]{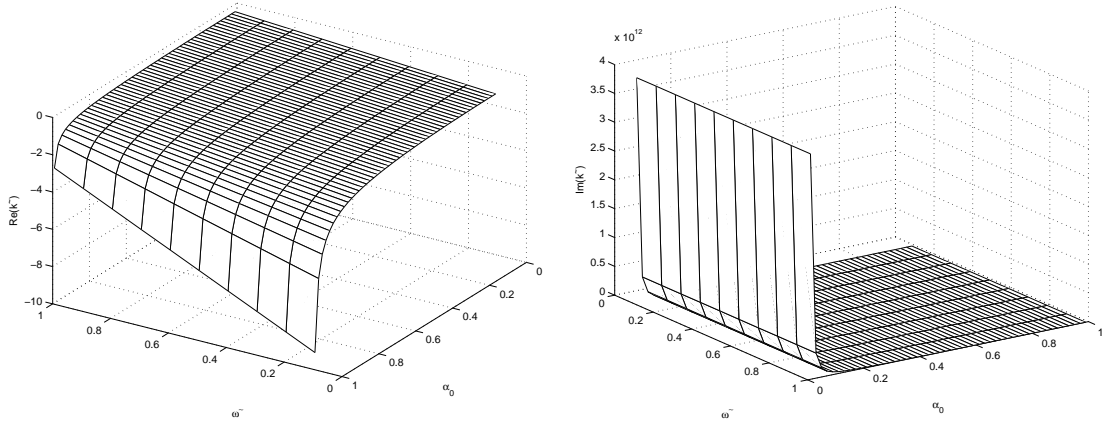}
\end{center}
\caption{\it Left: Real part of Alfv\'en  mode for the electron-ion
plasma. Right: Imaginary part of Alfv\'en damped mode.}
\end{figure}

\begin{figure}[h]\label{4}
\begin{center}
\includegraphics[scale=0.4]{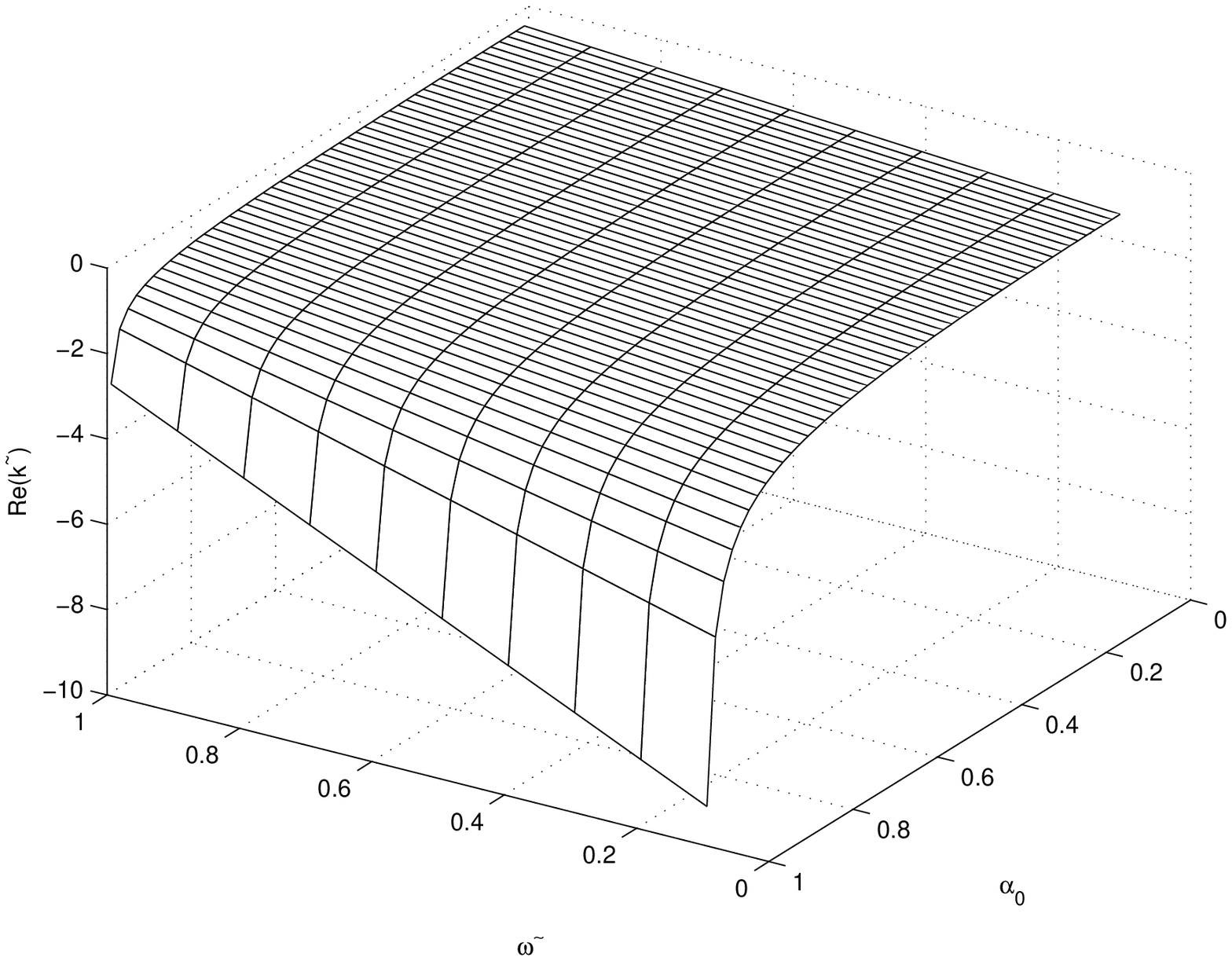}
\includegraphics[scale=0.4]{dtalfe_i_imag1_2.eps}
\end{center}
\caption{\it Left: Real part of Alfv\'en  mode for the electron-ion
plasma. Right: Imaginary part of Alfv\'en damped mode.}
\end{figure}

\begin{figure}[h]\label{5}
\begin{center}
\includegraphics[scale=.4]{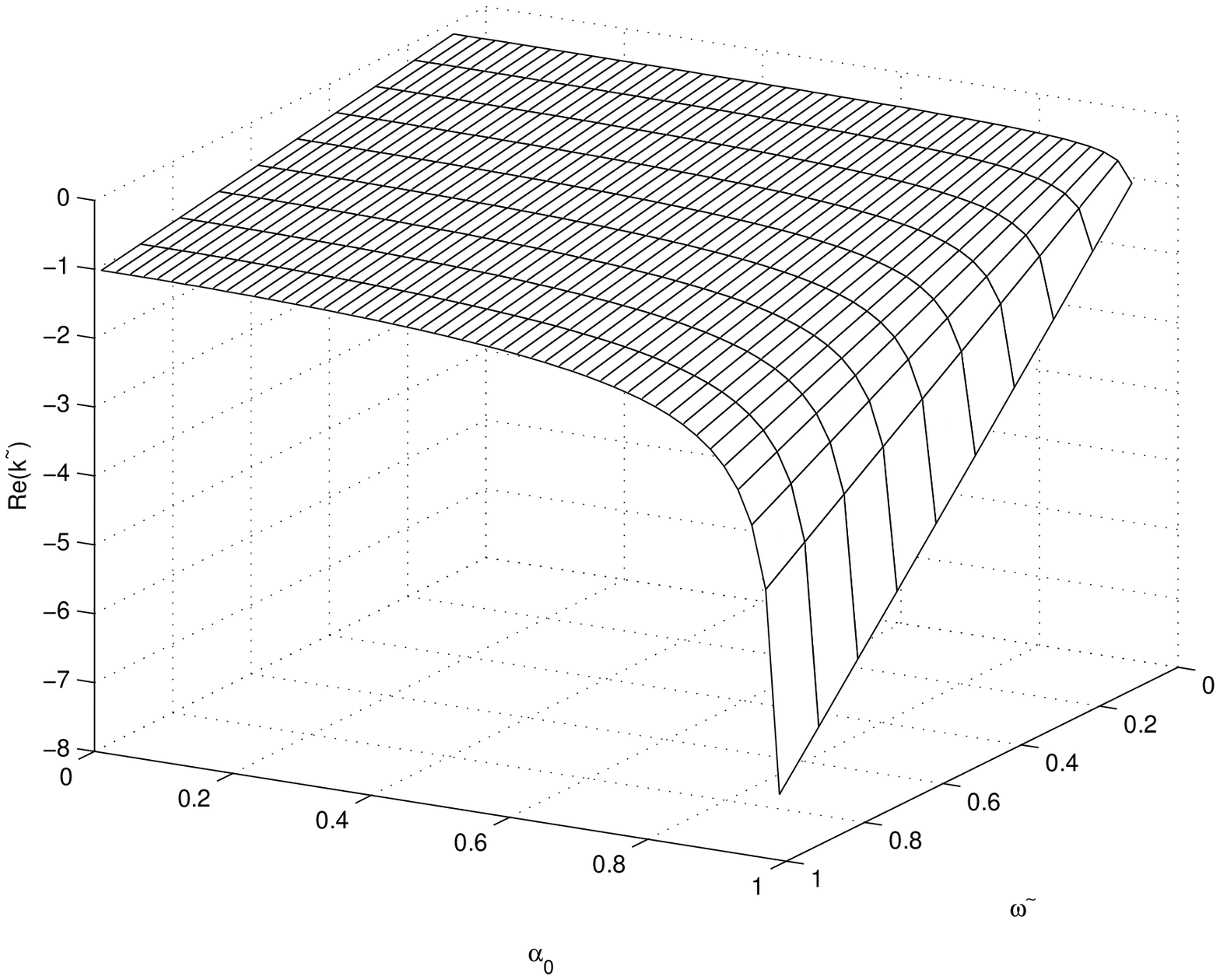}
\includegraphics[scale=.4]{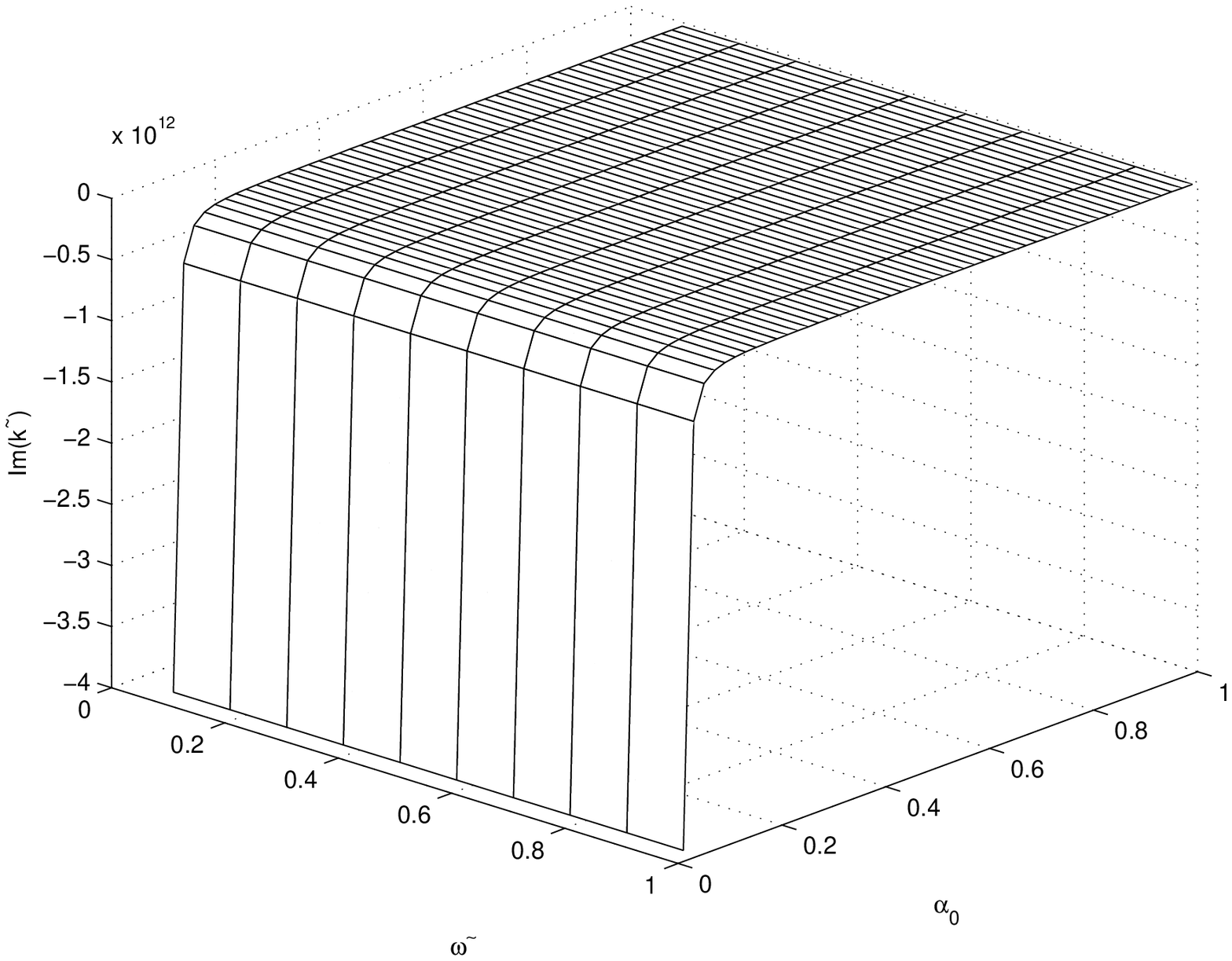}
\end{center}
\caption{\it Left: Real part of Alfv\'en  mode for the electron-ion
plasma. Right: Imaginary part of Alfv\'en  growth mode.}
\end{figure}

\begin{figure}[h]\label{6}
\begin{center}
\includegraphics[scale=.4]{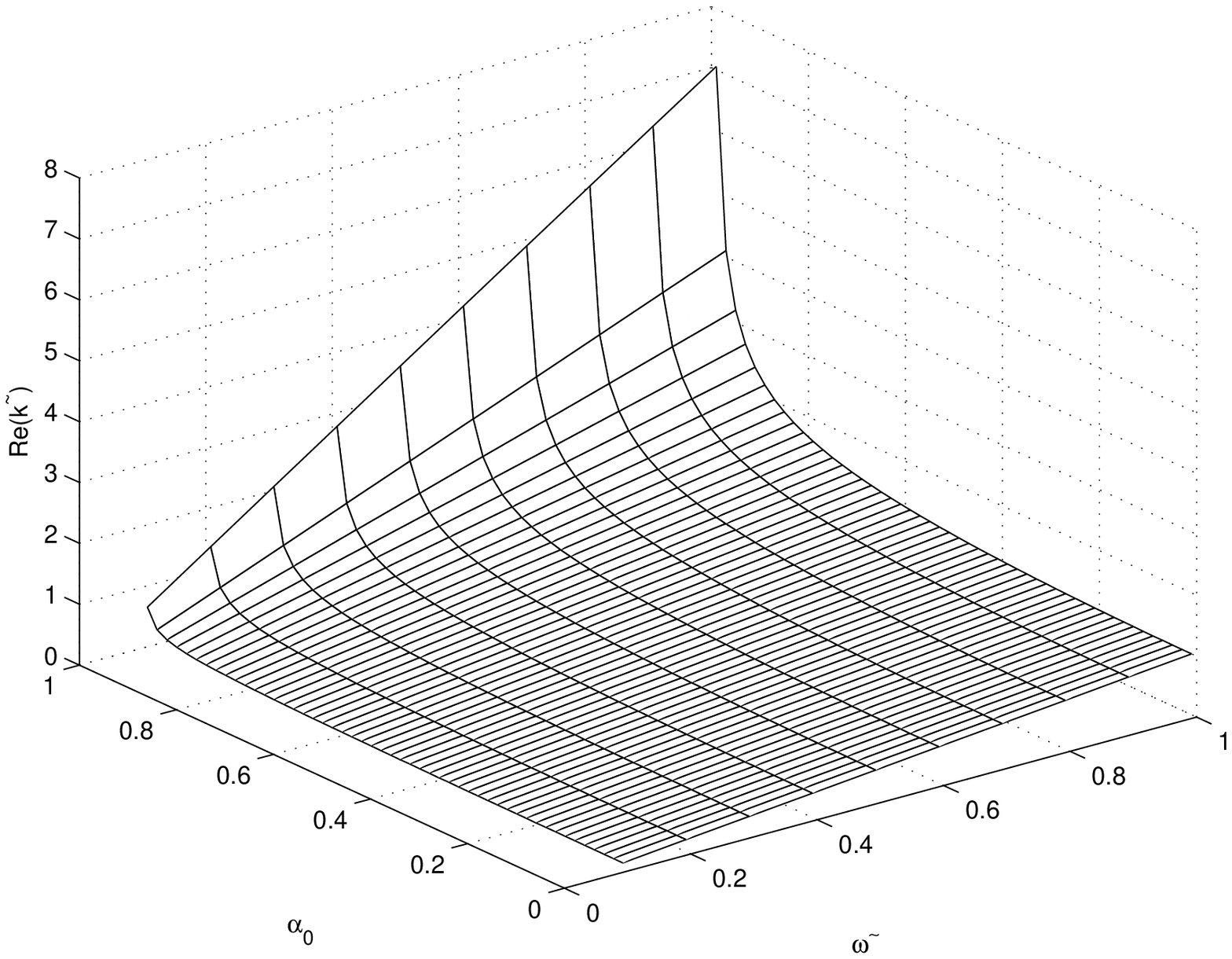}
\includegraphics[scale=.4]{dtalfe_i_imag3_4.eps}
\end{center}
\caption{\it Left: Real part of Alfv\'en  mode for the electron-ion
plasma. Right: Imaginary part of Alfv\'en  growth mode.}
\end{figure}

\begin{figure}[h]\label{7}
\begin{center}
\includegraphics[scale=0.4]{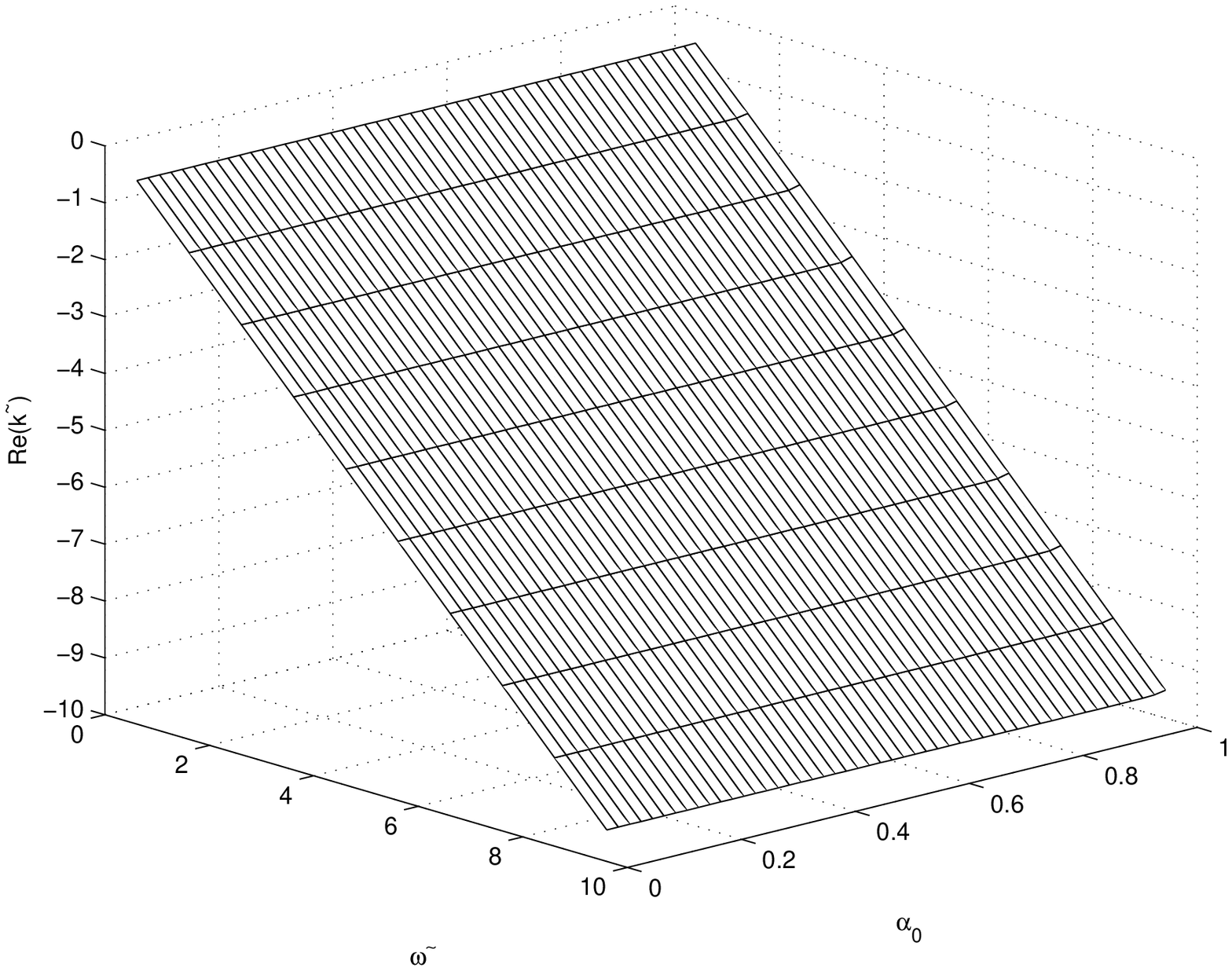}
\includegraphics[scale=0.4]{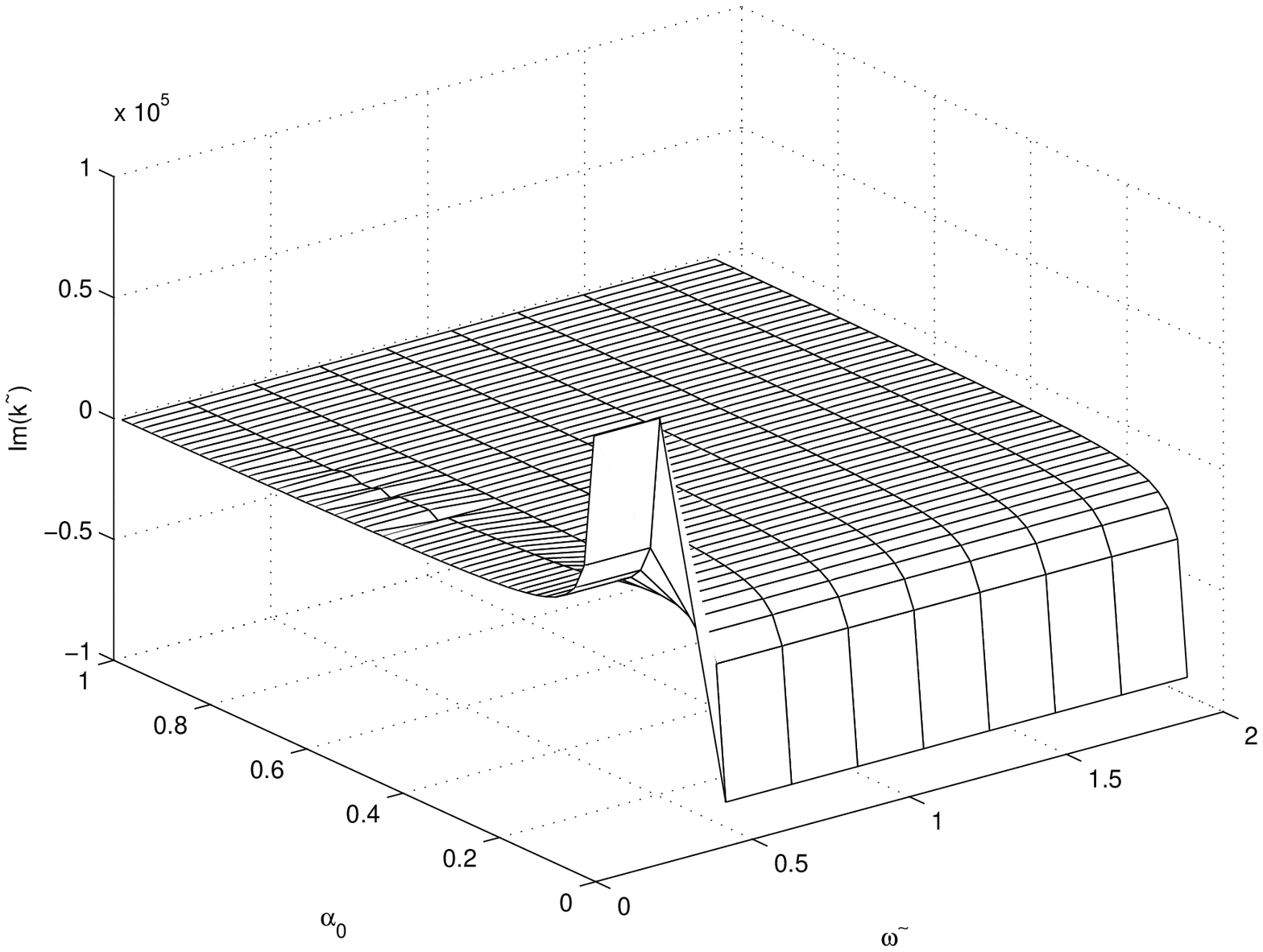}
\end{center}
\caption{\it Left: Real part of high frequency mode for the
electron-positron plasma. Right: Imaginary part of high frequency
damping and growth mode.}
\end{figure}

\begin{figure}[h]\label{8}
\begin{center}
\includegraphics[scale=0.4]{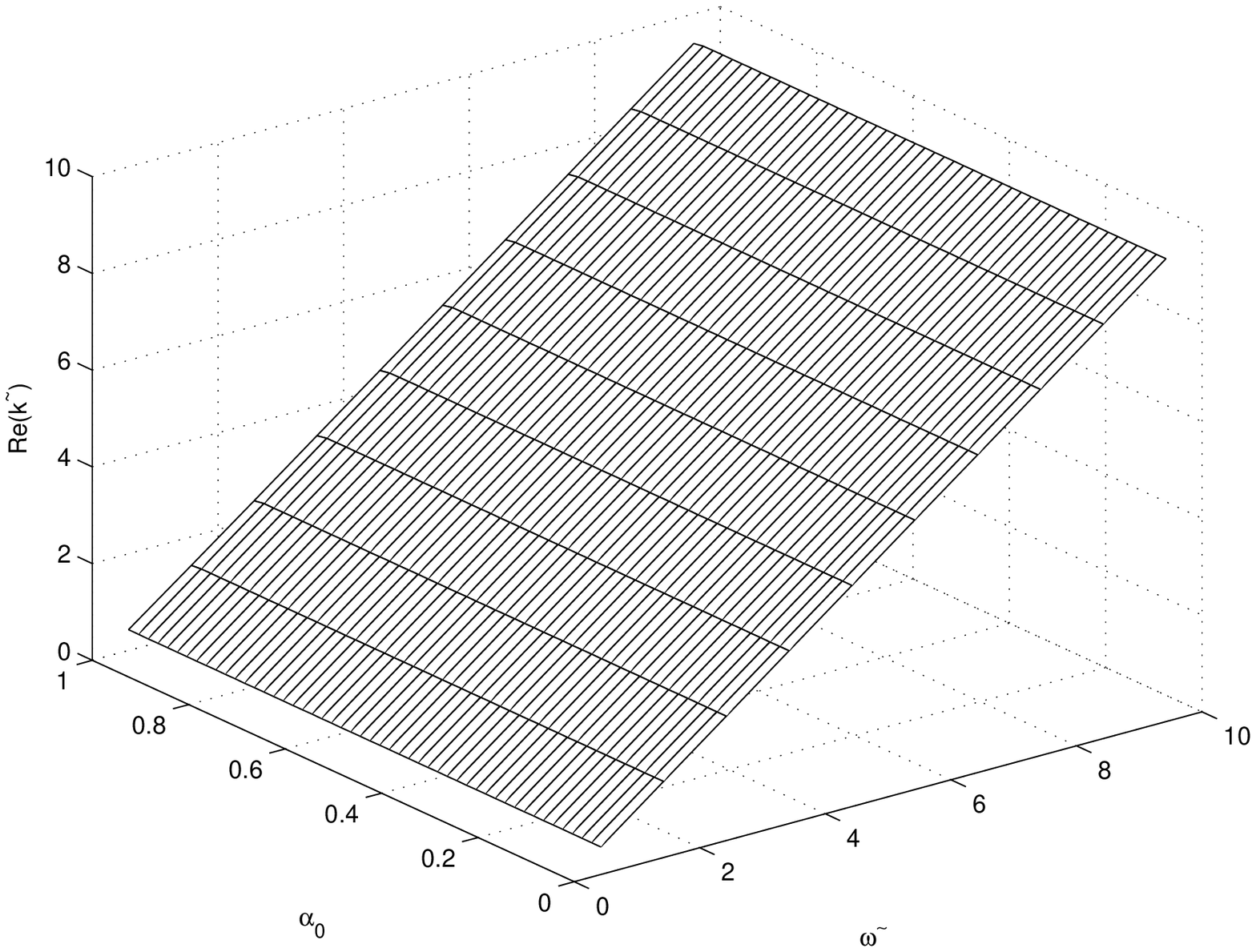}
\includegraphics[scale=0.4]{dthighe_p_imag1.eps}
\end{center}
\caption{\it Left: Real part of high frequency mode for the
electron-positron plasma. Right: Imaginary part of high frequency
damping and growth mode.}
\end{figure}

\begin{figure}[h]\label{9}
\begin{center}
\includegraphics[scale=.4]{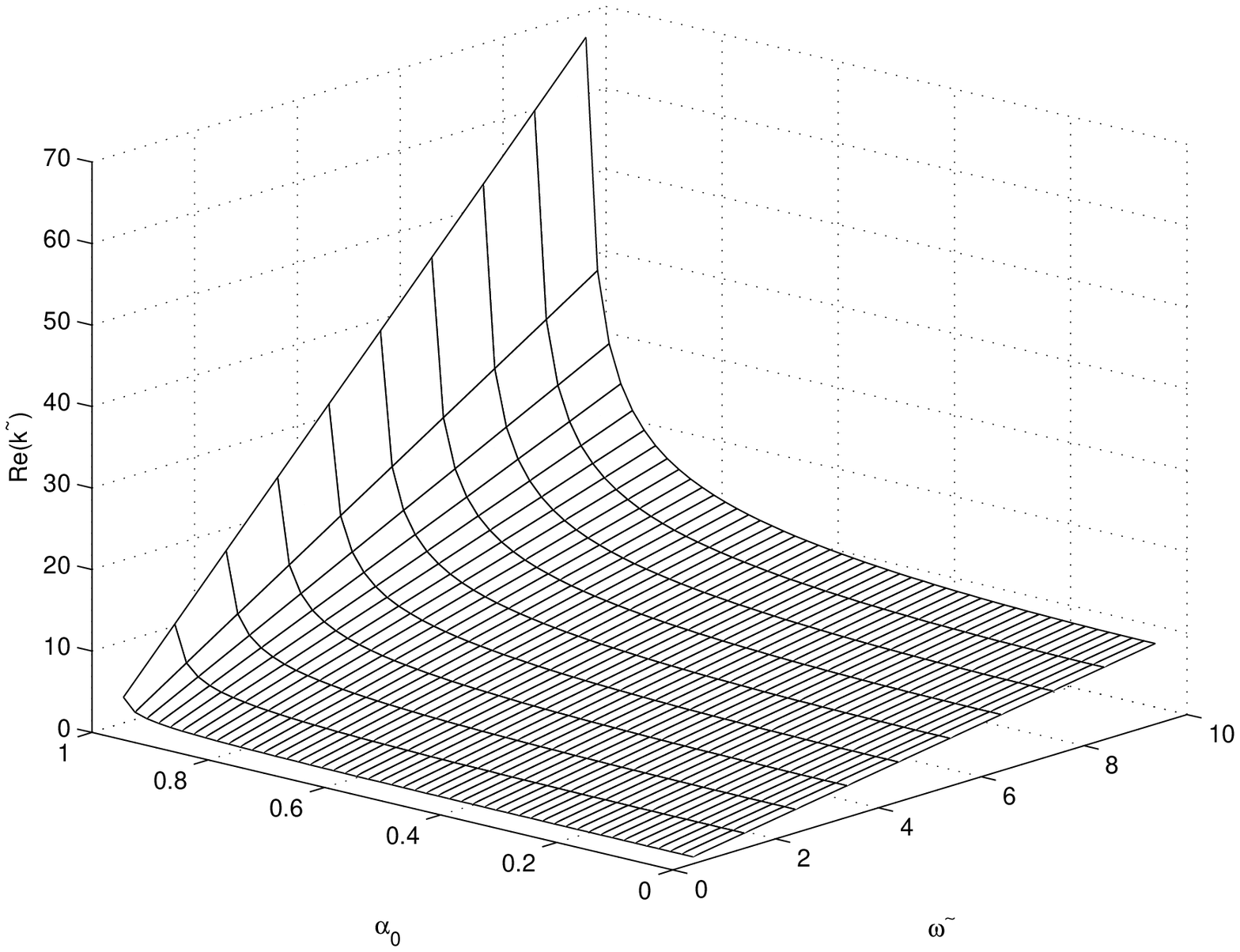}
\includegraphics[scale=.4]{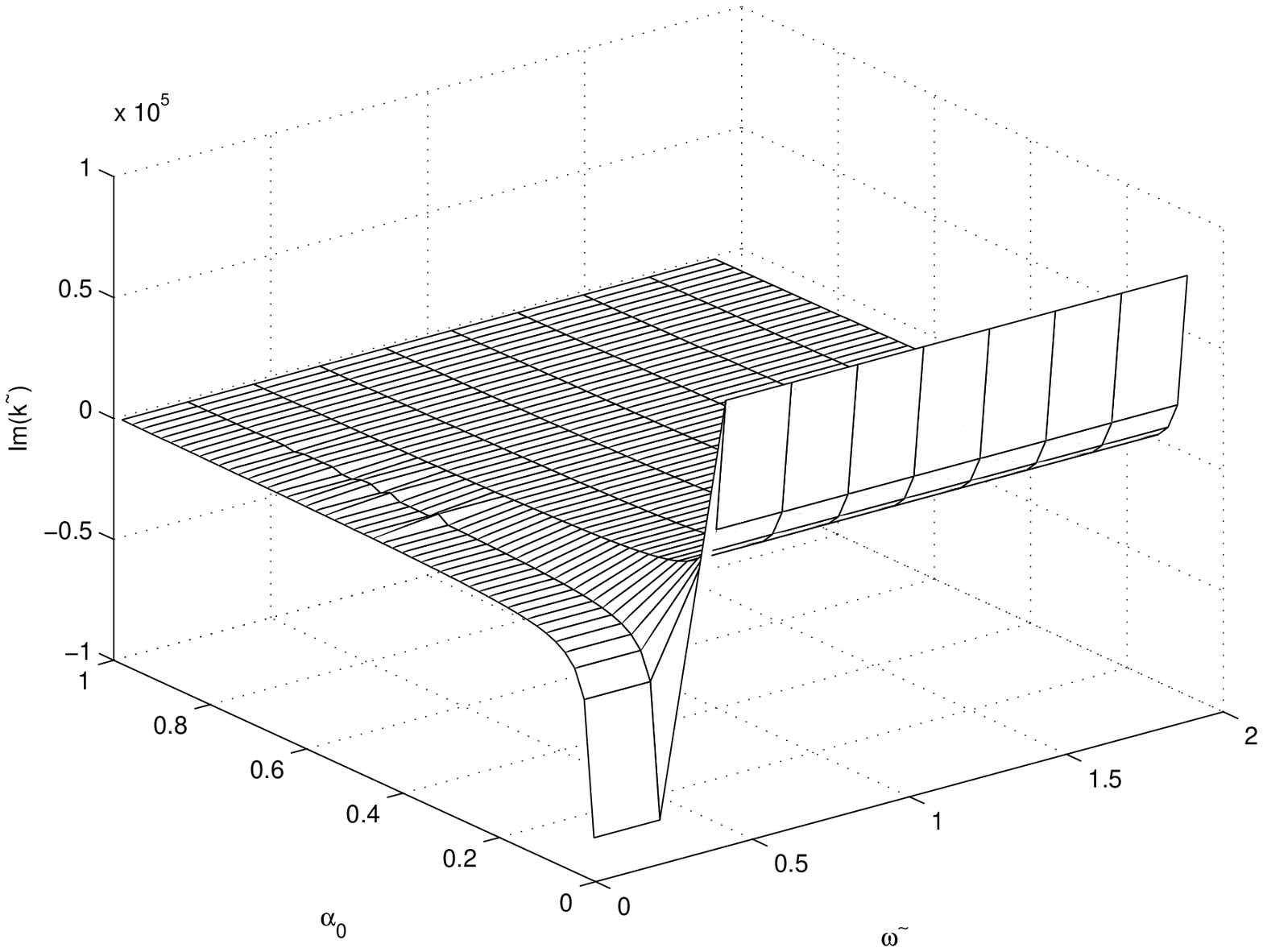}
\end{center}
\caption{\it Left: Real part of high frequency mode for the
electron-positron plasma. Right: Imaginary part of high frequency
growth and damping mode.}
\end{figure}

\begin{figure}[h]\label{10}
\begin{center}
\includegraphics[scale=0.4]{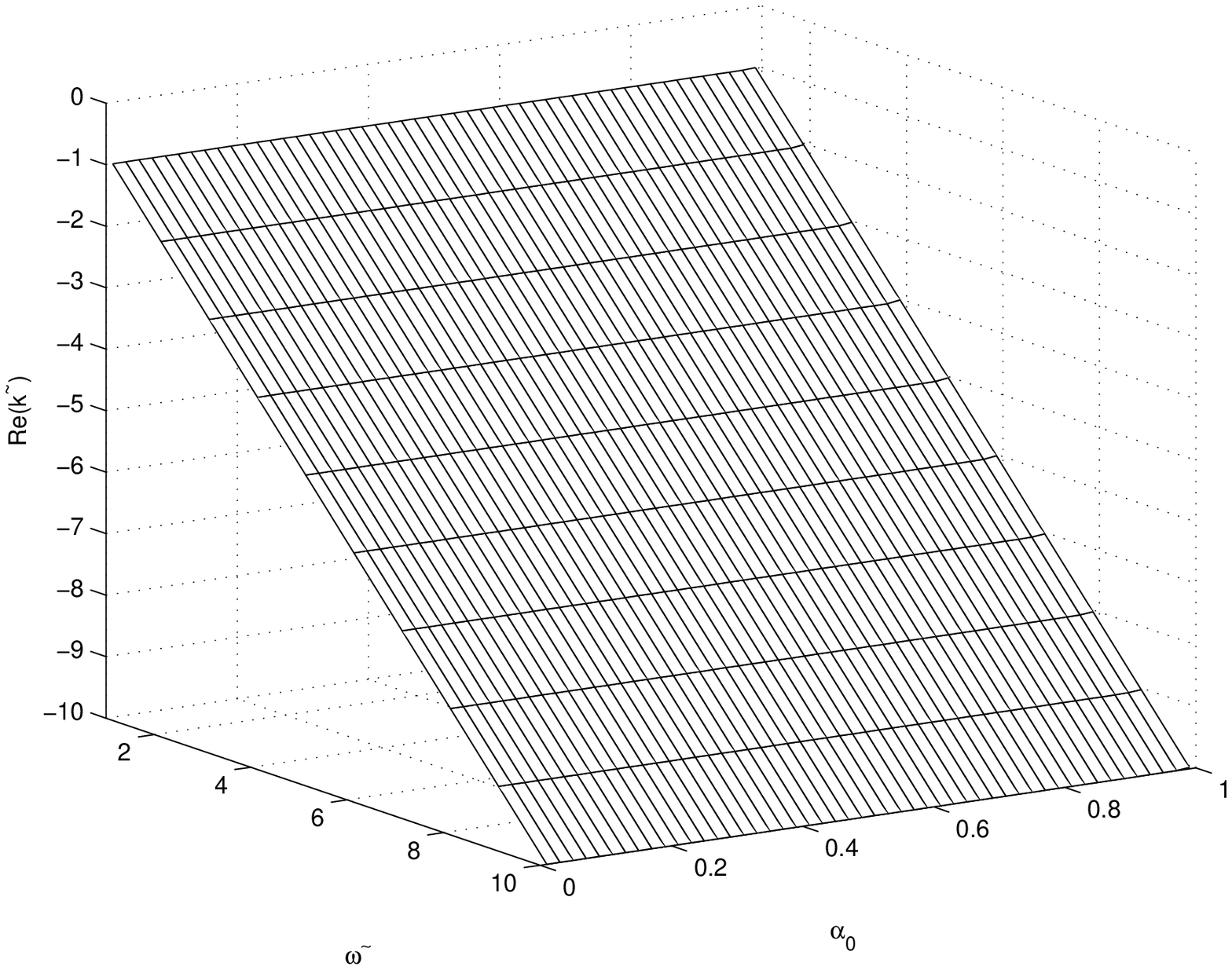}
\includegraphics[scale=0.4]{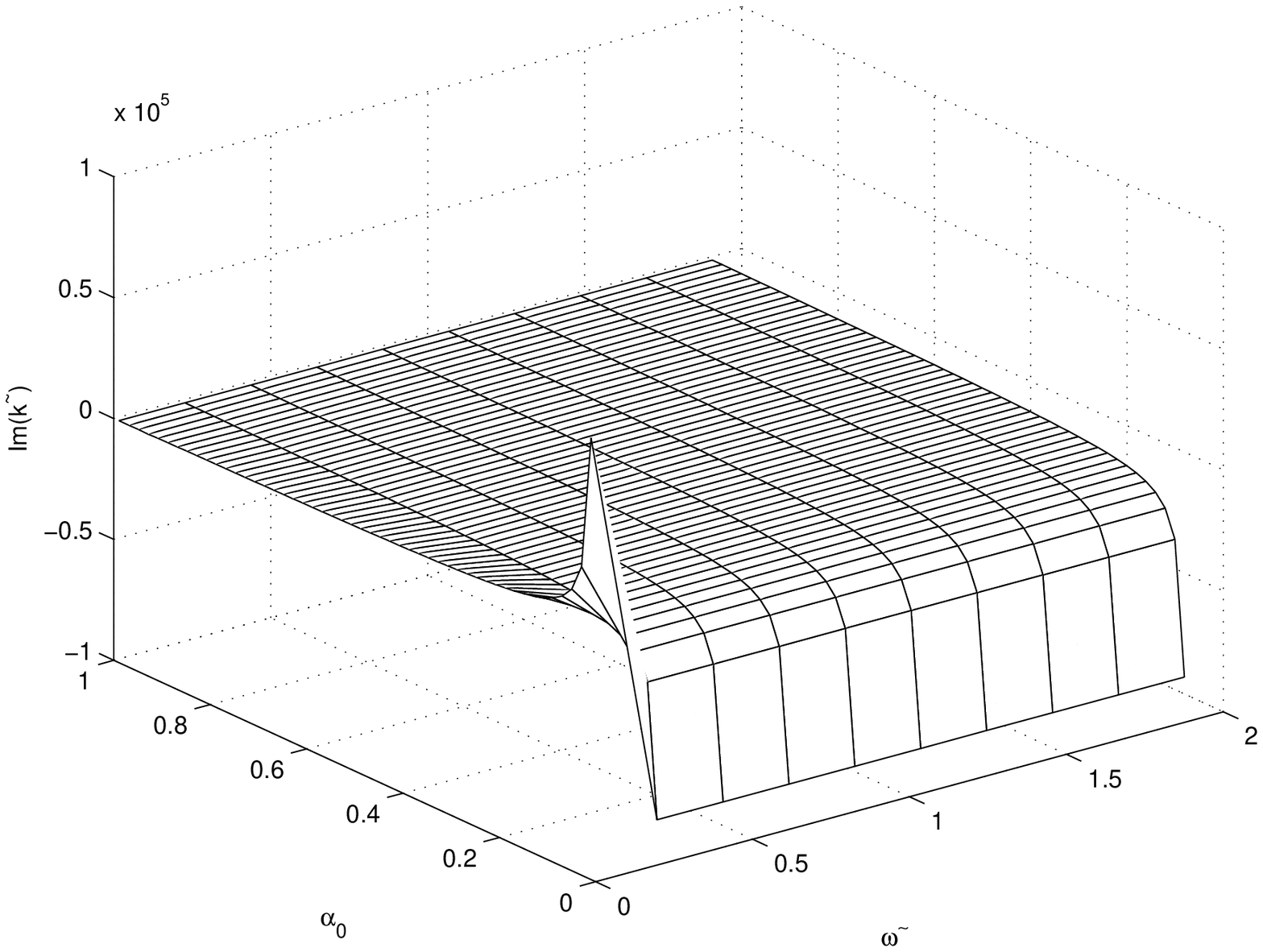}
\end{center}
\caption{\it Left: Real part of high frequency mode for the
electron-ion plasma. Right: Imaginary part of high frequency damping
and growth mode.}
\end{figure}

\begin{figure}[h]\label{11}
\begin{center}
\includegraphics[scale=0.4]{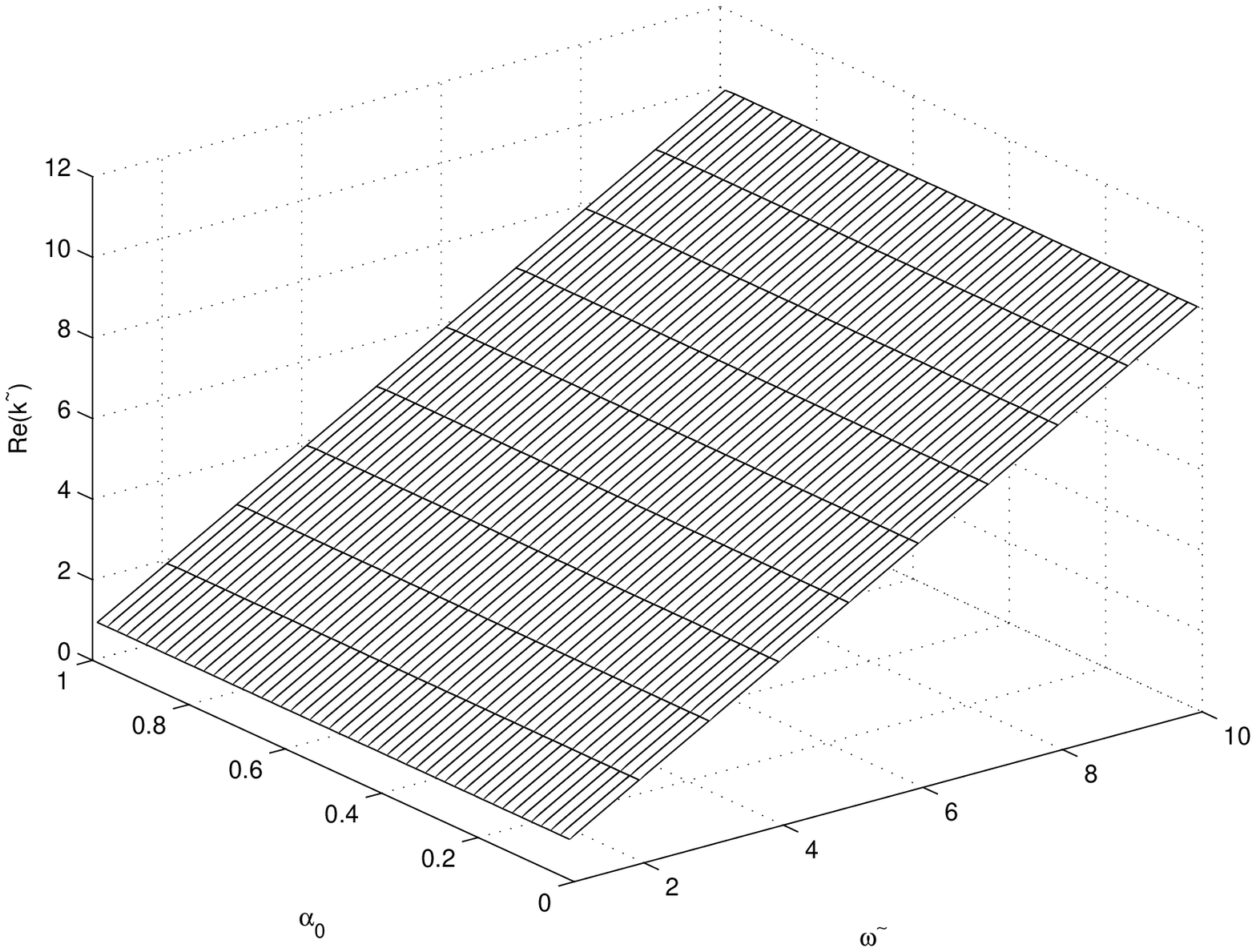}
\includegraphics[scale=0.4]{dthighe_i_imag1.eps}
\end{center}
\caption{\it Left: Real part of high frequency mode for the
electron-ion plasma. Right: Imaginary part of high frequency damping
and growth mode.}
\end{figure}

\begin{figure}[h]\label{12}
\begin{center}
\includegraphics[scale=.4]{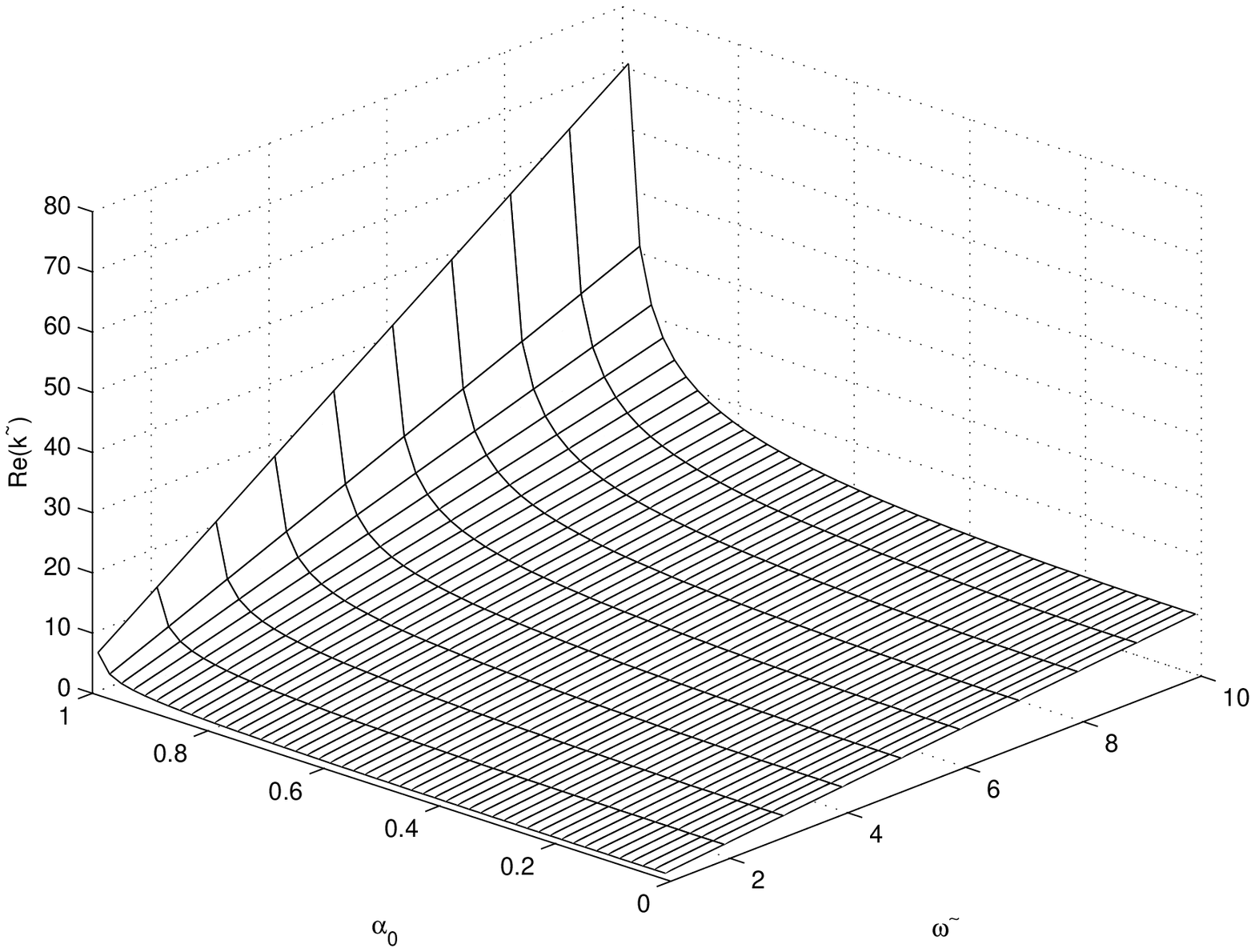}
\includegraphics[scale=.4]{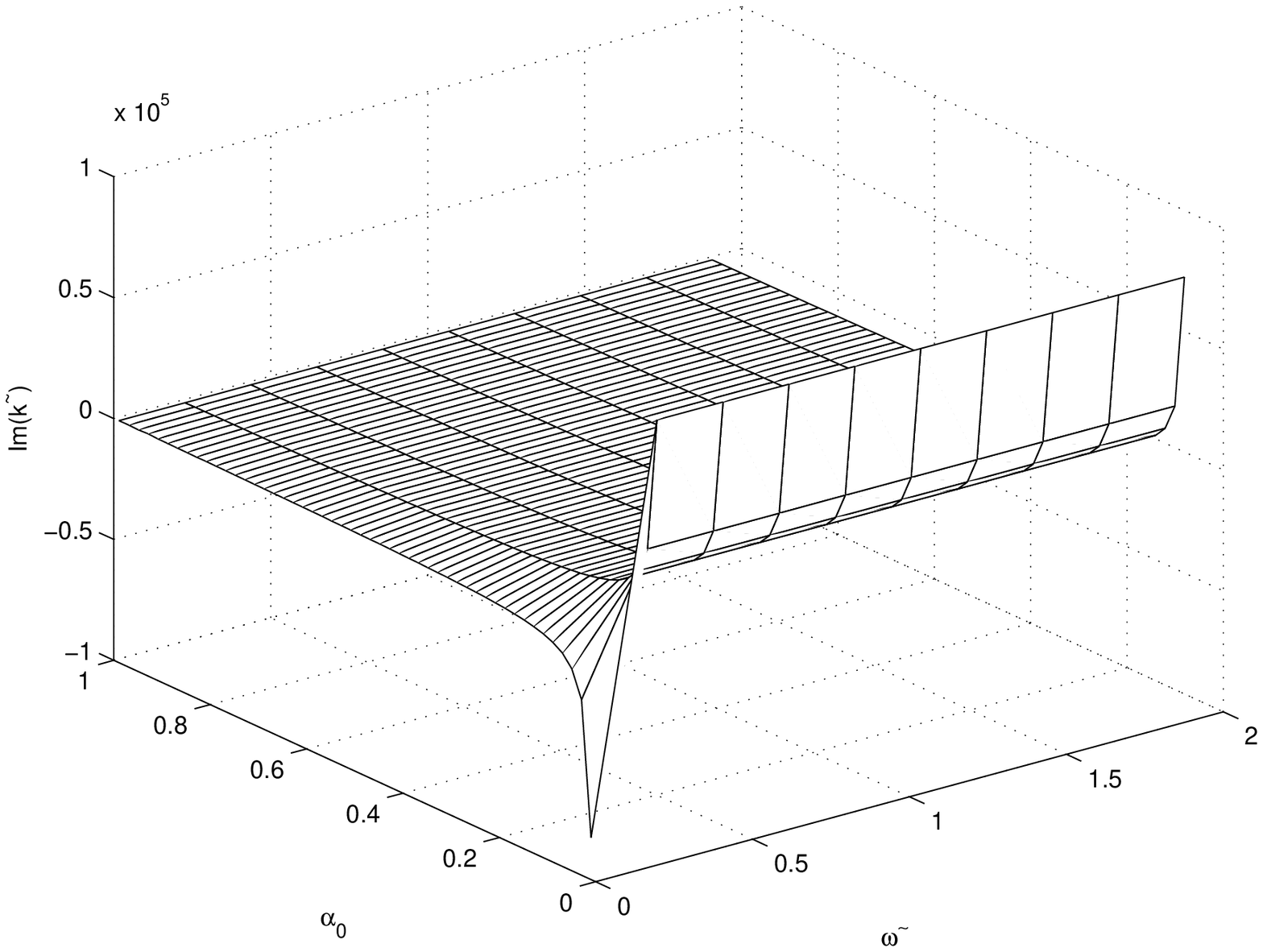}
\end{center}
\caption{\it Left: Real part of high frequency mode for the
electron-ion plasma. Right: Imaginary part of high frequency growth
and damping mode.}
\end{figure}

\end{document}